\documentclass[12pt]{iopart}

\usepackage{iopams}
\usepackage{graphicx}
\usepackage{color}
\usepackage[utf8]{inputenc}
\usepackage{subfiles}

\begin{document}

\review{Machine Learning for Condensed Matter Physics}

\author{Edwin A. Bedolla-Montiel$^1$, Luis Carlos Padierna$^{1}$ and Ramón Castañeda-Priego$^1$}

\address{$^1$ División de Ciencias e Ingenierías, Universidad de Guanajuato,
Loma del Bosque 103, 37150 León, Mexico}

\ead{lc.padierna@ugto.mx}

\begin{abstract}
    Condensed Matter Physics (CMP) seeks to understand the microscopic interactions of
    matter at the quantum and atomistic levels, and describes how these interactions result
    in both mesoscopic and macroscopic properties. CMP overlaps with many other important branches of science, such as Chemistry, Materials Science, Statistical Physics, and High-Performance Computing. With the advancements in modern Machine Learning (ML) technology, a keen interest in applying these algorithms to further CMP research has created a compelling new area of research at the intersection of both fields. In this review, we aim to explore the main areas within CMP, which have successfully applied ML techniques to further research, such as the description and use of ML schemes for potential energy surfaces, the characterization of topological phases of matter in lattice systems, the prediction of phase transitions in off-lattice and atomistic simulations, the interpretation of ML theories with physics-inspired frameworks and the enhancement of simulation methods with ML algorithms. We also discuss in detial the main challenges and drawbacks of using ML methods on CMP problems, as well as some perspectives for future developments.
\end{abstract}

\vspace{2pc}
\noindent{\it Keywords}: machine learning, condensed matter physics

\submitto{\JPCM}
\maketitle


\begin{figure}[htp]
    \centering
    \includegraphics[width=\textwidth]{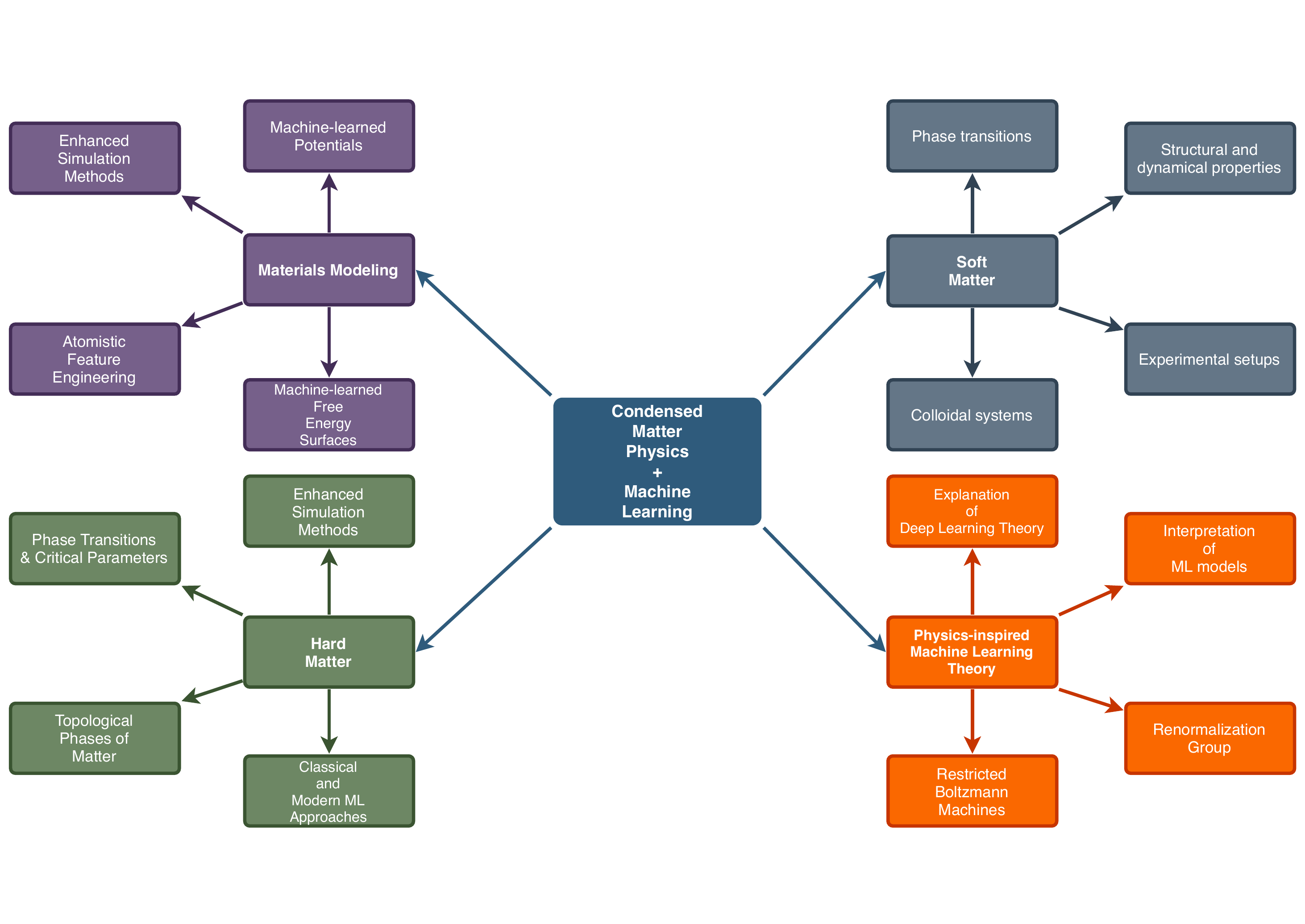}
    \caption{Schematic representation of the potential applications of ML to CMP discussed in this review.}
    \label{fig:schematic}
\end{figure}

\section{Introduction}
Currently, machine learning (ML) technology has seen widespread use in various aspects of modern society: automatic language
translation, movie recommendations, face recognition in social media, fraud detection,
and more everyday life activities~\cite{lecun2015deep} are all powered by a diverse application of ML methods.
Tasks like human-like image classification performed by machines~\cite{alom2018history};
the effectiveness of understanding what a person means when writing a sentence within a context~\cite{8416973};
the ability to distinguish a face from a car in a picture~\cite{voulodimos2018deep}; and automated medical image 
analysis~\cite{litjens2017survey} are some of the most outstanding advancements that stem from the mainstream use of modern ML
technology~\cite{Goodfellow-et-al-2016}.
Inspired by these notable applications, scientists have thrivingly applied ML technology to scientific fields,
such as matter engineering~\cite{Sanchez-Lengeling360}, drug discovery~\cite{vamathevan2019applications} and
protein structure predictions~\cite{senior2020improved}.

Influenced by the overwhelming effectiveness of ML technology in other scientific fields, physicists have also
applied such methods to specific subfields within Physics, with the main objective of making progress in the most challenging
research questions. The motivating reasons for this surge of interest at the intersection between Condensed Matter Physics (CMP) and
ML are vast. 
For instance, images, language and music exhibit power-law decaying correlations equivalent to those
seen in classical or quantum many-body systems at their critical points~\cite{carrasquilla2020machine};
very large data sets, with high-dimensional inputs found in materials science~\cite{schmidt2019recent} are the quintessential problem
that modern ML methods are tailored to deal with.
It seems as though ML is well-suited to be applied to CMP research problems given all these deep connections between both, but not without its drawbacks. Such is the case of
the encoding into a latent space of features that represent
the atomistic structure found in soft matter and physical chemistry molecular systems~\cite{Behler2016}, which
is a challenge that is currently being solved by modern natural language processing and object detection techniques. The challenge of encoding and feature selection, as well as other key challenges of ML applications to CMP will be discussed with further detail later on.

In this work, we review the contributions of ML methods to Physics with the aim of providing a starting point for both computer 
scientists and physicists interested in getting a better understanding of the interaction between these two research fields. 
For easier navigation of the topics covered in this review, the diagram shown in \fref{fig:schematic} displays a schematic
representation of the outline, but we shall summarize them briefly here.

We start the review with a very brief overview of both CMP and ML for the sake of self-containment, to help reduce the gap between both areas. In particular, the overview of CMP is meant to emphasize the difference between both main branches, i.e. Hard and Soft Matter, because these areas have a fundamental Physical difference that will later influence the way ML techniques are applied to their respective types of problems.

We then continue by exploring some of the research inquiries within \emph{hard condensed matter physics.}
\textbf{Hard Matter} has seen most of the ML applications, for instance in phase transition detection and critical phenomena for lattice models. We continue
with the review by analyzing how both classical ML and Deep Learning (DL) techniques have been applied to obtain detailed descriptions
of strongly correlated, frustrated and out-of-equilibrium systems.
We investigate how standard simulation methods, like Monte Carlo, have also seen enhancements from using ML models.

Moving forward, we examine a new research area dubbed \textbf{physics-inspired ML theory}, an area where the most rigorous and 
fundamental physics frameworks have been applied to ML to obtain a deeper insight of the learning mechanisms of ML models.
We discuss how in this engaging research area a robust understanding of the learning mechanism behind Restricted Boltzmann
Machines has been obtained. We move on to reviewing how common theoretical physics frameworks, such as the
Renormalization Group, have been used with interesting results to give insights
into the learning mechanisms of the most common models in DL.

The review continues with the exploration of \textbf{Materials modeling, Soft Matter, and related fields}, which recently have seen a strong surge
of ML applications, mainly in the subfields of materials science and colloidal systems,
e.g., proteins and complex molecular structures. 
Enhanced intelligent modeling, atomistic feature engineering, ML potential and free
energy surfaces
have been the primary topics of research; with the identification of a phase transition as well as the determination of the phases of matter being a close second.

As one has to be aware that every technique has some limitations, the review is closed with some of the main challenges and drawbacks of ML applications to CMP, in addition to some perspectives and outlooks about current challenges.

Before we continue with the review, we would like to address that many other notable applications within realated fields of CMP have been omitted due to lack of depth of knowledge in such fields and space in this short review.
Such applications include graph-based ML force fields~\cite{webb_graph-based_2019} and the deep learning architecture SchNet used to model atomistic systems~\cite{schutt_schnet_2018}; deep learning methodologies used to predict the ground-state energy
of an electron and effectively solving the Schr{\"o}dinger equation for various potentials~\cite{mills_deep_2017};
the use of ML techniques for many-body quantum states~\cite{carleo_solving_2017,glasser_neural-network_2018};
the use of crystal graph convolutional neural networks to directly learn material properties from the connection of atoms in the crystal~\cite{xie_crystal_2018};
the utilization of ML supervised and unsupervised techniques to approximate density functionals, as well as providing some insights into how these techniques might achieve such approximations\cite{snyder_finding_2012,li_understanding_2016}.
Lastly, we would like to suggest the reader interested in applications of ML to a wide range of areas within Physics, the recenet review by Carleo \etal~\cite{carleo_machine_2019} discusses in detail some of the most relevant works and research areas.

\section{Overview of Machine Learning}
\begin{figure}[htp]
    \centering
    \includegraphics[width=\textwidth]{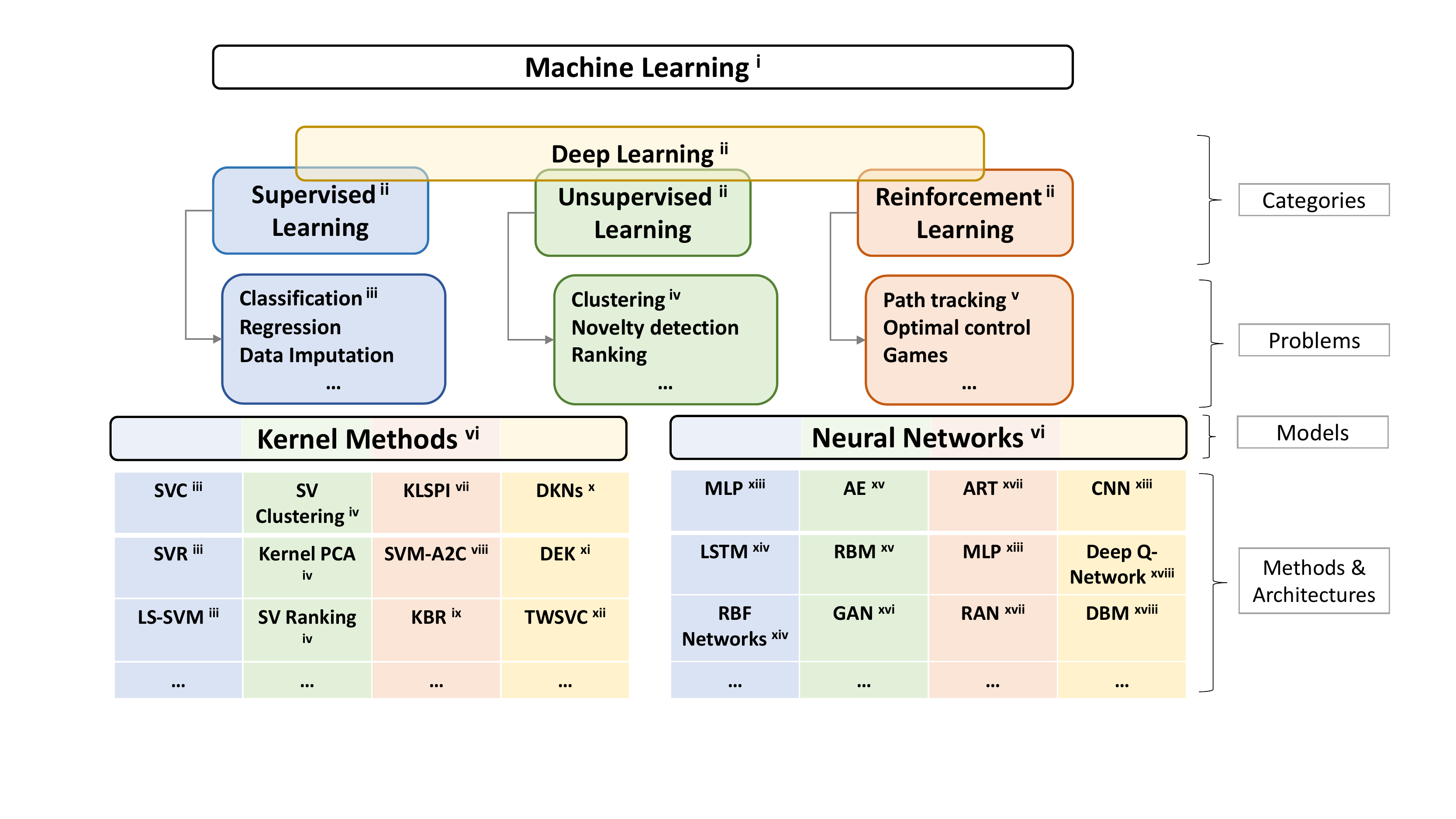}
    \caption{Partial taxonomy of Machine Learning methods and architectures. (i) A detailed description of other models not presented in the figure, such as graphical or Bayesian models can be consulted in~\cite{ethem2014a}. (ii) For further information regarding the ML categories please confer to~\cite{Goodfellow-et-al-2016} and~\cite{alom2019state}. (iii) Examples of methods for classification, regression and data imputation are discussed in~\cite{vapnik1998statistical},~\cite{smola2004tutorial}, and~\cite{zhang2009data}, respectively. (iv) Examples of methods for clustering, novelty detection and ranking can be found in~\cite{ben2001support},~\cite{hoffmann2007kernel}, and~\cite{tan2016dense}, respectively. (v) Examples of methods for path tracking, optimal control and games are introduced in~\cite{liu2020multi},~\cite{nguyen2020deep},~\cite{mnih2015human}, respectively. (vi) For further information regarding kernel methods and neural networks models please confer to~\cite{shawe2004kernel} and~\cite{aggarwal2018neural}, respectively. (vii) KLSPI - Kernel-based least squares policy iteration~\cite{xu2007kernel}. (viii) SVM-A2C - Advantage Actor-Critic with Support Vector Machine Classification~\cite{an2018discrete}. (ix) KBR - Kernel-based relational reinforcement Learning~\cite{driessens2006graph}. (x) DKN - Deep Kernel Networks~\cite{jiu2019deep}. (xi) DEK - Deep Embedding Kernel~\cite{le2019deep}. (xii) TWSVC - Twin Support Vector Clustering~\cite{bai2019clustering}. (xiii) Some architectures have been used in more than one category, for instance, variants of Convolutional Neural Networks (CNN) were employed both in supervised and unsupervised problems~\cite{sanakoyeu2018deep}, and Multilayer Perceptrons (MLP) have been used for supervised and reinforcement learning~\cite{shawe2004kernel,tesauro1992practical}. (xiv) LSTM - Long short-term memories and RBF - Radial Basis Function Networks~\cite{aggarwal2018neural}. (xv) AE - Autoencoders and RBM - Restricted Boltzmann Machines~\cite{alom2019state}. (xvi) GAN - Generative Adversarial Networks~\cite{liu2018unsupervised}. (xvii) ART - Adaptive Resonance Theory Networks~\cite{da2019survey} and RAN - Resource Allocation Networks~\cite{tan2008integrating}. (xviii) Deep Q-Network~\cite{mnih2015human} and DBM - Deep Boltzmann Machines~\cite{salakhutdinov2009deep}.}
    \label{fig:taxonomy}
\end{figure}

In this section, common ML concepts including those used in this review are described, following a specific hierarchy composed of category,
problem, model, and method or architecture. Figure \ref{fig:taxonomy} illustrates the relationship between these concepts. The interested reader can get further information from this partial taxonomy by following the references provided in the figure caption. 

Machine Learning is a research field within Artificial Intelligence aimed to construct programs that use example data or past experience to solve a given problem \cite{ethem2014a}. The ML algorithms can be divided into categories of supervised, unsupervised, reinforcement and deep learning \cite{bishop2006pattern,ethem2014a,Goodfellow-et-al-2016}: in supervised learning the aim is to learn a mapping from the input data to an output whose correct values are provided by a supervisor; in unsupervised learning, there is no such supervisor and the goal is to find the regularities or useful properties of the structure in the input; in reinforcement learning, the learner is an agent that takes actions in an environment and receives reward or penalty for its actions in trying to solve a problem, the output is a sequence of actions, and the main purpose is to learn from past good sequences to generate a policy that maximize a reward; deep learning is a special kind of ML, designed to deal with high-dimensional environments. DL may be regarded as a transverse category, since it has been used to solve the limitations of traditional algorithms of supervised \cite{hu2020deep}, unsupervised \cite{sanakoyeu2018deep}, and reinforcement learning \cite{nguyen2020deep}.
DL is focused on learning nested concepts, e.g., the image of a person, in terms of simpler concepts, such as the corners, contours or textures within the image~\cite{Goodfellow-et-al-2016}. 
The DL approach enhances previous ML models such as neural networks and kernel methods on several aspects, for instance by extending the capability of dealing with large-scale data, and providing new architectures, operators, and learning methods for the solution of problems with higher precision~\cite{mnih2015human,hinton2006reducing}. As a result, DL methods are able
to deal with huge amounts of heterogeneous information, as well as learning directly from raw data without the user-dependent 
preprocessing step of receiving the main and filtered features of raw data as valid training vectors (feature extraction) required by classical ML methods.

Although Figure \ref{fig:taxonomy} presents several ML concepts, there exist other important models like Decision Trees, Gaussian Processes, and methods such as boosting, LASSO, ridge regression, $k$-means, etc.~\cite{quinlan1986induction,williams2006gaussian,roth2004generalized}, that are not included in this brief overview in order to focus on those concepts that frequently appear in the rest of the manuscript. Specifically we will restrict ourselves to the description of Neural Networks (feed-forward, autoencoders, restricted Boltzmann machines and convolutional) and kernel methods (support vector machines). These methods and architectures are now described, with the exception of restricted Boltzmann machines that, because of their relevance, are presented in subsequent sections. Useful references are included for readers interested in obtaining a deeper insight.

Artificial Neural Networks (ANNs) are models of ML easily found in science and engineering applications. ANNs attempt to simulate the decision process of neurons of biological central nervous systems~\cite{graupe2019a}. Many ANNs are computational graphs~\cite{aggarwal2020linear,aggarwal2018neural,guresen2011definition}. In these graphs, the nodes compute activation 
functions (sigmoidal, radial basis, and any other Tauber-Wiener function~\cite{chen1995universal}) according to the input data 
provided by the connecting edges. The learning process consists in assigning weights to the edges and updating them according to a 
learning algorithm\textemdash such as backpropagation, Levenberg-Marquardt, Stochastic Gradient Descent, Natural Gradient Descent, 
etc.~\cite{shrestha2019a,bottou2012stochastic,pascanu2013revisiting}\textemdash. The tasks that ANNs can solve depend on the architecture that specifies how data is processed 
and stored. The storage of information can be either by using an external memory~\cite{graves2016hybrid,quan2019recurrent} or as an implicit memory codified into the weights of the network, being the latter the most common approach.  
The number of current ANNs architectures is immeasurable, but a project trying to track them can be consulted 
in Ref.~\cite{van2019a}.

Multilayer feed-forward Neural Networks (FFNNs) are one of the simplest ANN architectures, where neurons are organized in layers~\cite{rosenblatt1958a}. The default version of FFNNs assumes that all neurons in one layer are connected to those of the next layer. Neurons of the input layer receive the data to be processed and then pass it to the next layer, successive layers feed into one another in the forward direction from input to output. The key point is that the transfer of information between layers is controlled by nonlinear activation functions evaluated by the neurons. A neat illustration of this architecture and further details about its variants can be consulted in~\cite{aggarwal2018neural} sect 1.2.2.
Two examples of these networks are the Multilayer Perceptrons and  Radial Basis Function Networks.
When the learning process in FFNNs is supervised, in addition to the training data\textemdash$X \subset \Re^d$\textemdash provided to the input layer, expected output data\textemdash$Y \subset \Re$\textemdash
labeling each input vector is required to verify the performance of the network. A learning algorithm is then employed based on the 
network output until the best approximation to an underlying decision function, $f:X\rightarrow Y$, is achieved. FFNNs networks have 
been proved to be universal approximators to any Borel measurable function~\cite{hornik1989multilayer}. In addition to the function
approximation, supervised FFNNs may be used for classification, prediction, and control tasks~\cite{jain1996artificial}. FFNNs have also been applied to reinforcement learning problems~\cite{tesauro1992practical,mnih2015human,frommberger2010qualitative}.

Autoencoders (AEs) constitute a different type of NN architecture,
considered a special case of FFNNs, and designed to learn a latent representation of the training 
data in order to perform data compression or dimensionality reduction~\cite{bourlard1988auto,snoek2012nonparametric}. AEs consist of 
two parts: an encoder function\textemdash$g(\bi{x}),\bi{x} \in X$\textemdash, and a decoder that produces a 
reconstruction, 
$\bi{r}=f(g(\bi{x}))$. The main purpose of AEs is to generate an efficient encoding of the input data instead of 
learning to copy this data perfectly. Despite the success of ANNs to solve a vast number of applications in both science and 
engineering~\cite{doi:10.1002/minf.201700123,talwar2018autoimpute}, this model has been criticized due to the lack of providing 
explicit representations of the learned solutions~\cite{vellido1999neural,hsieh2011forecasting,bohanec2017explaining}. Since ANNs are considered black-box models, it becomes difficult to interpret their results, which is specially important in applications such as medicine, business or self-driving cars, where the reliance of the model must be guaranteed~\cite{bohanec2017explaining}. However, new techniques have been developed to increase the interpretation capabilities of NNs in order to extract new insights from complex physical, chemical and biological systems~\cite{montavon2018methods}.  

Convolutional Neural Networks (CNNs) are DL architectures specialized for processing data with a grid-like topology (time-series as 
one-dimensional data or images as two-dimensional grid of pixels) that use convolution in place of general matrix multiplication in at 
least one of their layers. The simplest CNN includes an input layer that receives the raw data; next, a convolutional layer filters local regions from the input layer's ouput. The use of convolution brings many benefits: provides a mean for working with inputs of variable size, it has the property of translation invariance, it is more efficient in terms of memory than dense matrix multiplication because of its property of parameter sharing, and the feature extraction is performed without human interference. The convolution is performed on two elements, the input data and a kernel whose values are assigned automatically by the learning algorithm, instead of being predefined by the user. After convolution, a pooling layer is then used to down-sample the spatial dimensions; lastly, a fully connected layer computes the class scores, and an output layer provides the final result~\cite{Goodfellow-et-al-2016}.

Support Vector Machines (SVMs) are explicit ML models that belong to the family of kernel methods and emerged as an alternative to ANNs~\cite{vapnik1998statistical}. Both ANNs 
and SVMs have the ability to 
generalize and may be designed to solve practically the same tasks (classification, regression, density estimations, clustering, among 
others)~\cite{PADIERNA2018211,rojas2017optimal}. However, while ANNs were inspired by the biological analogy to the brain, SVMs were 
inspired by statistical learning theory~\cite{vapnik1999overview}. Because of their solid theoretical basis,
SVMs can provide explicit solutions with measurable generalization capability by conducting structural risk minimization;
by solving a convex quadratic programming problem, the dreaded problem of local minima that permeates ANNs is avoided;
and finally, the model can represent the learned solution based on few parameters~\cite{maldonado2014alternative,xu2016novel}.
However, SVMs are limited in the size of the datasets that can handle since training a standard SVM has a time complexity between $O(N^2 )$ and $O(N^3 )$, with $N$ the number of input vectors; furthermore, the associated Gram matrix of size $N\times N$ must be allocated~\cite{tsang2005core,sadrfaridpour2019engineering}. This limitation is presented for general purpose SVM solvers like LIBSVM~\cite{changlibsvm}. Different approaches have been followed to deal with large-scale datasets. For instance, if the kernel function can be limited to the linear one (as in high-dimensional sparce spaces), then efficient solvers like LIBLINEAR~\cite{fan2008liblinear} or PEGASOS~\cite{shalev2011pegasos} can be implemented to solve problems with hundreds of thousands of input vectors in seconds. Other strategies like sampling, boosting or hierarchical training have been used to speed up the SVM training with non-linear kernels~\cite{sadrfaridpour2019engineering, nandan2014fast}. For Deep Kernel Learning methods, where a concatenation of several kernel functions can be reformulated as neural networks~\cite{bohn2019representer}, some strategies focus in designing maps in the Reproducing Kernel Hilbert Space to reduce the complexity of evaluating Deep Kernel Networks, which scale quadratically on $N$ and linearly w.r.t the depth of the trained networks~\cite{jiu2019deep}. Based on these limitations and proposed solutions, an application-based analysis should be done to choose the correct set of SVM algorithms or related kernel methods.

\section{Overview of Condensed Matter Physics}
Understanding the intricate phenomena of all types of matter and their properties is the driving force of
condensed matter physicists. From quantum theory all the way to materials science, CMP encompasses a large
diversity of subfields within physics, as it deals with different time and length scales depending on both
the molecular details and the type
of matter being analyzed. To study such systems, theoretical, computational and experimental physicists
collaborate to gain a deeper insight into the behavior in and out of equilibrium of matter. We shall briefly talk about the two
main areas within CMP, namely,  \emph{hard matter} and \emph{soft matter}, the essential models as well as the
central problems these research areas are focused on. More information about these research areas can
be found in standard textbooks, see, e.g., ~\cite{chaikin1995principles,anderson2018basic,girvin_yang_2019}
and reviews~\cite{RevModPhys.71.S59,LUBENSKY1997187}.

\subsection{Soft Condensed Matter}
Fluids, liquid crystals, polymers, colloids, and many more are the principal types of matter that
are studied by Soft Condensed Matter, or Soft Matter (SM). SM has the peculiar property that it will
easily respond to external forces. SM will change its flow when a small shear
force is applied to it, or thermal fluctuations are applied, thus changing its properties as a whole.
We are surrounded by these materials in our everyday life, from glass windows, toothpaste, hair-styling
gel, among others. These materials can show diverse properties and when some external force is applied,
they will exhibit a different set of properties, but in all cases SM shows the fascinating property of
self-organization of its molecules~\cite{RevModPhys.71.S367}.
Proteins and bio-molecules are also good examples of SM, they change their structure when subjected to
thermal fluctuations, while also showing a pattern of self-assembly~\cite{Yan1882,ZHANG2002865},
as is the case in most types of SM.
All in all, SM is a subfield of CMP that gained a lot of attention when first introduced in the 1970s
by Nobel prize laureate Pierre-Gilles de Gennes~\cite{B419223K}, because it has shown to be of great importance
to science in general, with very useful applications, as well as pushing the boundaries in
theoretical, experimental and computational Physics.

In order to explore all these properties and changes in a SM system, a model that describes most of these types
of matter is needed. Most fluids and colloidal dispersions show a repulsive behavior, which can be modeled very well with
the \emph{hard-sphere} model~\cite{russel1991colloidal}.
A system modeled as a hard-sphere dispersion only shows an infinite repulsive interaction
when there exists interparticle separations less than the particle diameter, and zero otherwise. There is no attractive
interaction included in this model. The hard-sphere model is a very simple one, which has also exhibited intriguing, but
unexpected equilibrium and non-equilibrium states~\cite{PhysRevLett.106.105704,eberle2012dynamical,valadez2012phase}.
There are some interesting properties that define the hard-sphere model in SM. First of all, the internal energy
of any allowed system configuration is always zero. Furthermore, forces between particles and variations in the
free energy\textemdash defined from classical thermodynamics as $F = E - TS$, with $E$ being the internal energy,
$T$ the temperature and $S$ the entropy\textemdash are both determined entirely by the entropy.
Moreover, the entropy in a hard-sphere system depends directly on the total volume occupied by the hard spheres,
commonly defined as the \emph{volume fraction}, $\phi$.
When $\phi$ is small then the system shows the properties of an ideal gas, but as $\phi$ starts to increase,
particles interact with each other and their motion is restricted by collisions with other nearby particles.
Even more interesting is the fact that the \emph{phase transitions} of hard-sphere systems are completely defined
by $\phi$~\cite{cheng2001phase}. In general, a SM system does not need to be composed only of hard spheres, it can
also be built with non-spherical molecules, for example, rods, sphero-cylinders\textemdash cylinders with semi-circular 
caps\textemdash, and
hard disks, just to mention a few examples. Nevertheless, all these systems experience a large
diversity of phase transitions~\cite{doi:10.1080/00268979809483199,McGrother_1996,Bleil_2006}.

The description of phase transitions and critical phenomena are one of the main challenges in all CMP. When a thermodynamic system changes its uniform
physical properties to another set of properties, then the system encounters a phase transition. In SM,
phase transitions are one of the most important phenomena that can be analyzed given the diversity of phases
that could appear in unexpected circumstances~\cite{anderson2002insights, zinn2007phase}.
In particular, a gas-liquid phase transition is defined by a temperature\textemdash, known as the
\emph{critical temperature}, $T_c$\textemdash, and a density\textemdash, called the \emph{critical density},
$\rho_c$\textemdash. These quantities are also known as the \emph{critical point} of the system, meaning that
two phases can coexist when the system is held to these special values.
When neither of these thermodynamical quantities can be employed, an \emph{order parameter} is
more than sufficient to determine the phase transitions of a system. Order parameters are special quantities that
can measure the degree of order between a phase transition; they range between several values for the different phases
in the system~\cite{binney1992theory}.
For instance, when studying the liquid-gas transition in a fluid, a specified order parameter can take the value of
zero in the gaseous phase, but it will take a non-zero value in the liquid state. In this scenario, the density difference
between both phases $\left(\rho_l - \rho_g\right)$\textemdash where $\rho_l$ is the density in the liquid state and $\rho_g$ in the
gas state\textemdash can be a useful order parameter.

When traversing the phase diagram of a SM system, one might encounter the so-called \emph{topological defects}.
A topological defect in ordered systems happens when the \emph{order parameter space} is continuous everywhere except at isolated points, lines or surfaces~\cite{RevModPhys.51.591}. This can be the case of a system which is stuck in-between two phases, and there is no continuous
distortion or movement from the particles that can return the system to an ordered
phase.
One such example is that of nematic liquid crystals~\cite{PhysRevE.47.3343}. The nematic phase is observed in a hard-rod system, when
all the rods align in a certain direction. Hard rods are allowed to move and to rotate freely.
Furthermore, the neighboring hard rods tend to align in the same direction. But if it so happens that some hard rods are aligned in
one direction and the rest in another direction, then the system contains topological defects~\cite{GILVILLEGAS1997441}.
These defects depend on the symmetry of the order parameter of the phases, as well as the topological properties of the space in which
the transformation is happening. Topological defects are of the utmost importance in a large variety of systems in SM, e.g., they
determine the strength of the liquid crystal mentioned before.

Nevertheless, these types of systems and problems are not so easily solved. How can we tell when a system will be prone to topological defects?
Is it possible to determine the phases that a system will come across given its configuration? Even more challenging questions
are always originating when these types of exotic matter appear.
But theoretical, experimental or even computational frameworks are not enough to tackle these problems.
Physicists are always open to new ways to approach such challenges, and ML is slowly starting to become a powerful tool to deal with these complex systems,
and in this review we shall discuss some of the ways ML has been applied to such problems, as well as commenting on some of the problems that can arise from using such techniques.

However, it is important to note that SM also experiences thermodynamic transitions to solid-like phases.
For example, in the case of the hard-sphere model, when the packing fraction $\phi$ approaches the value of $\phi \approx 0.492$~\cite{alder_phase_1957,fernandez_equilibrium_2012}, it undergoes a liquid-solid first order transition~\cite{hoover_melting_1968,robles_note_2014}.
This thermodynamic behavior can also be studied using theoretical and experimental techniques employed when one deals with Hard Matter.
Thus, at some point, SM is also at the boundary Hard Matter.

\subsection{Hard Condensed Matter}
When physicists deal with materials that show structural rigidity, such as solids, metals, insulators or semiconductors,
their interest in the properties of these materials are the main focus of Hard Condensed Matter, or Hard Matter (HM).
Furthermore, in contrast to SM systems, HM typically deals with systems on smaller scales, i.e., matter that is governed by
atomic and quantum-mechanical interactions. Entropy no longer determines the free energy of a HM system,
it is now up to the internal energy.
One particular example of interest in HM is the phenomenon of \emph{superconductivity}.
A material is considered a type I superconductor if all electrical resistance is absent and it is a perfect diamagnet~\cite{annett_2004}
\textemdash i.e., all magnetic flux fields are excluded from the system\textemdash; a type II superconductor is not completely diamagnetic because they do not exhibit a complete Meissner effect~\cite{tinkham_1996}.
The phenomenon of superconductivity is described by both quantum
and statistical mechanical interactions. This means that superconductivity can be seen as a macroscopic manifestation of the laws
of quantum mechanics. As was the case with SM, we are also surrounded by HM materials, for instance, semiconductors are the
principal components in microelectronics~\cite{SANTI2018157}, which in turn are the components of every digital device we use,
ranging from computers to smart cellphones.
The applications that stem from HM are too great to enumerate in this short review, and the theoretical,
experimental and computational frameworks that have derived from HM are nothing short of revolutionary; we suggest the reader who might be interested in knowing some of these applications to a technical perspective article~\cite{yeh2008perspective}, and a more thorough book on some of the modern aspects of HM, as well as its applications to technology~\cite{NAP11967}.

In HM, we also need a reference model that can help explain at the basic level some of the investigated materials. Many other materials can use a similar reference model, but with different assumptions and physical quantities.
Such is the case of lattice systems, and, in particular, the Ising model~\cite{doi:10.1080/00029890.1987.12000742}.
The Ising model is the simplest model that can explain ferromagnetic properties, in addition to accounting for quantum interactions.
It is defined in a lattice, much like if it were a crystalline solid.
In each site of the lattice, there is a discrete variable
$\sigma_k$ that takes one of either two values, such that $\sigma_k \in \{-1, +1\}$.
The energy for a lattice is given by the following Hamiltonian

\begin{equation}
    H(\sigma) = -\sum_{\langle i, j \rangle} J_{ij} \sigma_i \sigma_j - \mu \sum_{j} h_j \sigma_j ,
    \label{eq:ising}
\end{equation}

where $J_{ij}$ is a pairwise interaction between lattice sites; the notation $\langle i, j \rangle$ indicates that $i$ and $j$ are
lattice site neighbors; $\mu$ is the magnetic moment and $h_j$ is an external magnetic field. As originally solved by Ising in
1924~\cite{ising1925contribution}, when the system lies in a one-dimensional lattice, the system shows no phase transition.
Later, Lars Onsager proved in 1944~\cite{onsager1944} that when the Ising model lies in a two-dimensional lattice a continuous
phase transition is observed. For dimensions larger than two, different theoretical~\cite{PhysRevB.4.3174,PhysRevB.4.3184}
and computational~\cite{PhysRevB.44.5081} frameworks have been proposed.

While the Ising model has seen some very useful applications, it is not the only model that has shaped the research interests within
CMP, as it cannot possible contain and explain all phenomena seen in HM. A generalized model, such as the $n$-vector
model~\cite{PhysRevLett.20.589} or the Potts model~\cite{PhysRev.64.178} are extensively studied models in HM that have
been successfully applied to explain even more intricate phenomena~\cite{PhysRevB.46.662,Straley_1973}.

In general, all these models show phase transitions, in the same way
as SM systems. This research area grew even more when the Berezinsky-Kosterlitz-Thouless (BKT) phase transition was
discovered~\cite{Berezinsky:1970fr,Berezinsky:1972rfj,Kosterlitz_1973,Kosterlitz_1974}. The BKT transition was unveiled on
two-dimensional lattice systems, such as the XY model\textemdash which is a special case of the $n$-vector model\textemdash.
The BKT transition consists of a phase transition that is driven by topological excitations and does not break any symmetries
in the system~\cite{Chen_2017}. This type of phase transition was quite abstract in the 1970s when first introduced, but it turned
out to be a great theoretical framework that was later discovered in exotic 
materials~\cite{PhysRevB.87.184505,PhysRevLett.109.017002}, which in turn showed puzzling properties.
Such was the impact of the BKT transition in CMP that it would later be awarded a Nobel prize in
Physics in 2016~\cite{gibney2016physics}.

However, the BKT transition is a complicated transition to study. Hard condensed matter physicists are always trying to find a better
way to look for these kind of unique aspects in materials. It is an elusive and complicated challenge.
In the same spirit as in SM,
ML has already been put to the test on this task, if it is possible to re-formulate the problem at hand as a ML problem,
it can then be solved by standard ML techniques. But, as mentioned before, the task of re-formulation and feature selection is not simple; there is currently no standard way to approach this in CMP applications, contrary to most ML models and applications.
We shall examine these applications and their potential challenges as well as drawbacks with more detail in the following sections.

\section{Hard Matter and Machine Learning}
In this section, we shall review the main techniques and ML models applied to CMP, mainly to HM.
These applications might constitute a paradigm of ML applications to CMP due to the results obtained, which are generally acceptable.
The structure of HM data is similar to the one found in computer vision applications in ML, among other applications.
Furthermore, these schemes have also shown most of the major shortcomings such as feature selection, training and tuning of ML models.

\subsection{Phase transitions and critical parameters}
One of the most prominent uses of NNs was on the two-dimensional 
ferromagnetic Ising model which has a Hamiltonian similar to Eq.~\eref{eq:ising},
with the aim to identify the critical temperature, $T_c$, at which the system leaves an ordered 
phase\textemdash the \emph{ferromagnetic phase}\textemdash and enters a disordered
phase\textemdash the \emph{paramagnetic phase}. Carrasquilla and Melko~\cite{Carrasquilla2017}
introduced a new paradigm by essentially formulating the problem of phase transitions as a \emph{supervised classification}
problem, where the ferromagnetic phase constitutes one class and the paramagnetic phase is another one, then
a ML algorithm is used to discriminate between both of them under specific conditions.
By sampling configurations using standard Monte Carlo simulations for different system sizes,
and then feeding this raw data to a FFNN, an estimate of the 
correlation-length critical exponent $\nu$ is obtained directly from the output of the
last layer of the trained NN.
Finite-size scaling theory~\cite{doi:10.1063/1.330232} is then employed to obtain an estimation
of $T_c.$ Even if the geometry of the lattice is modified, the power of
abstraction provided by the NN~\cite{aggarwal2018neural} is enough to predict the critical
parameters even if it was trained in a completely different geometry; this makes it a very convenient tool
to identify critical parameters for systems that do not have a defined order parameter~\cite{kogut1979}.

\subsection{Topological phases of matter}
Once having shown that ML is able to identify phases of matter in simple models as
shown for models that have a similar Hamiltonian to Eq.~\eref{eq:ising},
researchers have been interested in trying out these algorithms in more complex systems, and 
a lot of work has been put into the topic of identifying topological phases of matter using ML
algorithms for strongly correlated topological systems~\cite{PhysRevB.96.195145,broecker2017machine,tanaka2017detection,PhysRevB.96.245119,rodriguez2019identifying,PhysRevB.97.045207,Melko_2004,PhysRevX.7.031038,PhysRevB.96.205146,10.5555/3122009.3242020,PhysRevB.99.075113}.
For instance, Zhang \etal~\cite{Zhang2019} have applied FFNNs as well as CNNs to find the BKT phase transition, as well as
constructing a complete phase diagram of the XY and generalized XY models, with very promising results.
In the case of the FFNN approach,
Zhang \etal found out that a simple NN architecture consisting of one input layer, one hidden layer with four
neurons and one output layer is more than enough to obtain the critical exponent $\nu$ for the XY model.
We can compare this to the results obtained by Kim and Kim~\cite{PhysRevE.98.022138}, where they explain a similar
behavior but for the Ising model. In their work, analyzing a tractable
model of a NN, they found out that a NN architecture consisting of one input layer, one hidden layer with two
neurons and one output layer is sufficient to obtain the critical exponent $\nu$ for the Ising model. If such is the case for a shallow architecture to be able to obtain good approximations for the critical exponent $\nu$ then, perhaps, such models can be further simplified into more robust and analytically tractable ones. Such is the case of the Support Vector Machine methodology developed by Giannetti \etal~\cite{Giannetti2019}.

It is important to keep track of the model complexity specially for complicated models to avoid overfitting
and to reduce the amount of data needed to train the model.
In the case of the NN methodologies exposed before, we can argue that such models are easier to train if there are enough data samples in the data set, when this is not the case then overfitting will occur yielding low accuracy results; NNs are also faster to train due to the fact that these models can be implemented in accelerated computing platforms, such as Graphical Processing Units (GPUs).
On the other hand, SVMs cannot be readily accelerated with such techniques but they are less prone to overfitting because the model complexity, in cases where the \emph{support vectors}\textemdash the subset of the data set needed to construct the \emph{decision function} and thus make a prediction\textemdash are less than the total number of samples, is low enough that a smaller data set might be enough to obtain desirable approximations. Still, SVMs need to be adjusted according to the task, so hyperparameter tuning must always be part of the data processing pipeline, which in turn might take longer to obtain results.
If this is the case, then simpler methods\textemdash such as logistic regression or kernel ridge regression~\cite{friedman2001elements}\textemdash could potentially be better suited for the task of approximating critical exponents due to the fact that these models are simple to implement, need less data samples and most implementations are numerically stable and robust;
at the very least these simple models could potentially provide a starting point for such approximations.

New and interesting methodologies are always arising in order to solve the problem of phase transitions.
Kenta \etal~\cite{kenta2020machine} developed a generalized version
of the methodology proposed by Carrasquilla and Melko~\cite{Carrasquilla2017} for a variety of models, namely, the Ising model,
the $q$-state Potts model~\cite{RevModPhys.54.235} and the $q$-state clock model.
Instead of using the raw data obtained from Monte Carlo simulations as proposed in~\cite{Carrasquilla2017},
Kenta \etal considered the configuration of a specific long-range spatial correlation function for each model,
effectively exploiting feature engineering, hence providing meaningful data to the NN, which now contains relevant information
from the systems at hand. This strategy has the advantage of generalization for the already trained NN, thus it can be readily
applied to similar systems without further manipulation of the physical model or the production of new simulation data. The strategy
is also quite general in the sense that it can be applied to multi-component systems,
with the similar precision as standard theoretical methods.

Finally, we remark some of the advances in topological electronic phases.  The most fundamental quantity  to  characterize  these  states  is  the  so  called  topological invariant, whose value determines the topological class of the system. 
In particular, interfaces between systems with different topological invariants show topologically  protected  excitations,  resilient  towards  perturbations respecting the symmetry class of the system. Mappings of topological invariants using NN as performed in~\cite{PhysRevB.97.115453} shows that supervised learning is an efficient methodology to characterize the local topology of a system.
DL has also seen some useful applications to these type of systems, e.g., for random quantum systems~\cite{ohtuski2016,ohtuski2017,ohtuski2020}.

\subsection{Classical and modern ML approaches}
The amount of success that ML has seen in the area of Lattice Field Theory and Spin Systems has
given it a special place in the toolboxes of physicists in order to provide a different perspective to a given problem based on different ML models, each with its own advantages and drawbacks.
It is specially important to identify those tools as the ones that have seen the most use in this topic,
the main one being FFNNs given its generalization and abstraction capabilities for which
it has become the workhorse of most of the applications in this field\textemdash particularly
when the problem can be formulated as a \emph{supervised learning} problem.

For more classical approaches, SVMs have also been implemented on the phase transition
problem~\cite{PhysRevB.96.205146,Giannetti2019} with promising results given their ease of use
and physical interpretation of the results, which we shall discuss in detail in~\sref{interpretation}. It is useful to compare the SVM method against the NN one presented in the previous subsection, specially given the fact that both techniques can yield very similar results; equivalently we can ask ourselves, when should we use one technique or the other?
As mentioned in the previous subsection, SVMs are a great tool when the number of samples in the data set is low, this is because SVMs only need a small subset, i.e., the support vectors, to effectively predict some value. The main disadvantage of SVMs is their time complexity when training, as it completely depends on the implementation used and the data set. For example, the most popular implementation of a SVM solver is LIBSVM~\cite{changlibsvm}, which scales between $\Or (M \times N^2)$ and $\Or (M \times N^3)$, where $M$ is the number of \emph{features} and $N$ is the number of samples in the data set. SVMs depend strongly on the hyperparameters used, so in the case of an incorrectly tuned SVM the time it takes to train could be $\Or (N^3)$ in the worst-case scenario, while also yielding poor results. On the other hand, if the user is careful in tuning and selecting a good kernel as presented in~\cite{Giannetti2019}, one can solve a CMP problem with very high precision; in this case some Physics insight of the problem can be helpful in providing good parameters for the SVM.

If we now turn our attention to NNs and their training time complexity, it normally scales as $\Or (N^3)$ provided the NN use an efficient linear algebra and matrix multiplication implementation. If we also account for the backpropagation operations, which scale linearly with the number of layers in the network, we can see that NNs are significantly more difficult to train than SVMs. Research is being carried out to understand why NN are hard to train for  and at the same time they are efficiently trained in practice~\cite{livni_computational_2014,song_complexity_2017}.
The reason for this is that modern numerical linear algebra implementations are efficient, and they can be accelerated with modern computing technologies, such as GPUs. This makes NNs a great tool when the number of data samples is large, while training can also happen \emph{online}, i.e. feeding the network with newly created samples. In short, both techniques give similar results provided there are enough representative data samples from the problem.
A simple criterion for choosing between both methodologies depends greatly on the data set and the problem, but starting off with SVMs can give a good approximation, and in the case of needing even more precision or if the data set contains a large number of features then NNs might be a better choice for the problem at hand.

Many other classical techniques\textemdash specifically the ones that stem from \emph{unsupervised learning} methods
\textemdash have become prominent algorithms. For instance, one such technique is 
the nonlinear dimensionality reduction algorithm
t-distributed Stochastic Neighbor Embedding (t-SNE)~\cite{maaten2008visualizing}
that Zhang \etal~\cite{Zhang2019} used to map high-dimensional features onto lower dimensional spaces
in order to identify three distinct phases in the site percolation model. In this methodology, the idea is to use different site configurations, handled as a single array of values from the lattice, and using t-SNE to map this large dimensional space into a 2-dimensional one, such that when enough iterations are performed three distinct clusters appear: one for each ordered phase (percolating and non-percolating phases), and the transition phase between these two phases.
Although t-SNE is mostly a visualization tool, by employing other unsupervised learning techniques such as $k$-means~\cite{friedman2001elements} or spectral clustering~\cite{Ng01onspectral}, together with the results obtained from t-SNE one could possibly obtain a good approximation of the transition point in the percolation model.

Another prime example is Principal Component Analysis (PCA)~\cite{doi:10.1080/14786440109462720,friedman2001elements}
which has proved to be one of the best feature extraction algorithms that can be used in Spin Systems~\cite{PhysRevE.96.022140,PhysRevB.94.195105},
as it can be readily applied to raw configurations and use the information from the system and build different features, which then can be compared to physical observables like structure factors and order parameters.
For instance, Wei~\cite{PhysRevB.94.195105} established a methodology to build a matrix $M$with different site configurations from the Ising model\textemdash using a simplified version of the Hamiltonian from Eq.~\eref{eq:ising}\textemdash and then performing PCA on $M$. This yields a set of new \emph{component vectors} that describe the large dimensional space of $M$ in a new, lower dimensional space. These new vectors are used as a basis set to project the original samples in this space and it is in this new space that clustering techniques like $k$-means can be used in order to find a phase transition.

DL has also been wielded as a powerful technique to understand quantum phase
transitions~\cite{PhysRevB.97.134109} in the form of adversarial NNs~\cite{goodfellow2014generative};
to provide a thorough description of many-body quantum systems~\cite{rocchetto2018learning}
by employing Variational Autoencoders~\cite{kingma2013auto,Goodfellow-et-al-2016};
and CNNs~\cite{Goodfellow-et-al-2016} have been applied, for instance, to learn
quantum topological invariants~\cite{PhysRevLett.120.066401}, or to solve frustrated many-body
models~\cite{PhysRevB.98.104426}. One of the main reasons for DL to be such a strong
asset is that it can handle complex data representations and achieve good accuracy results, provided these models have enough data to learn from. The main drawbacks of these models is precisely the fact that the data needed must be representative of the task at hand; the data must also preserve invariants of the system in the case that the system has such invariants, such as invariance to translation and rotations. These type of issues are mostly solved with \emph{data augmentation} techniques, such as adding some white noise to the data samples, or creating some additional samples that are mirrored, although there are other such modifications that can be done. But again, this implies that the user of such models must know the problem well enough to perform such adjustments for the sake of stable training and yielding good results.

\subsection{Enhanced Simulation Methods}
One of the standard computer simulation technique in HM, along with SM, is the Monte Carlo (MC) method, an unbiased probabilistic method
that enables the sampling of configurations from a classical or quantum many-body
system~\cite{baumgartner2012monte,newman1999monte,gubernatis2003monte,landau2014guide}.
Throughout the decades, specific modifications have been performed to the MC method in order to enhance it, creating
a better, more rigorous method that could be used in all types of simulations\textemdash
from observing phase transitions to computing quantum path integrals\textemdash.

It is in this situation that ML has something
to offer in order to improve the original MC method; such is the case of the Self-learning Monte Carlo (SLMC)
method developed by Li \etal~\cite{Liu2017}. In this novel scheme, the classical MC method is used to obtain
configurations from a similar Hamiltonian to Eq.~\eref{eq:ising}\textemdash these configurations will serve as
the training data\textemdash; we call this set of configurations $\{s\}$ from a state space $\mathcal{S}$ where each sample should have a probability density given by some function $p\colon \mathcal{S} \to \Re^{+}$.
Then a ML method is employed\textemdash linear regression, in this case\textemdash
to learn the rules that guide the configuration updates in the simulation, i.e. a parametrized effective model $\tilde{w}_{\theta}$ which can approximate the original model and obtain $\tilde{w}_{\theta} \approx w$, with $\tilde{w}_{\theta}\colon \tilde{\mathcal{S}} \to \Re^{+}$ and $\tilde{\mathcal{S}} \supseteq \mathcal{S}$.
This enables the MC method to \emph{learn by itself} under which conditions a given configuration is updated
without external manipulation or guidance, efficiently sampling configurations for the most general possible model.
Demonstrated on a generalized version of the Hamiltonian given by Eq.~\eref{eq:ising}, the SLMC method is extremely efficient
when producing configurations from the system, lowering the autocorrelation time from each of the updates.
Li \etal discussed that, although a ten-fold speedup from the original MC method is obtained, the SLMC still suffers from
the quintessential problem of the original MC method\textemdash when approaching the thermodynamical limit, 
the method heavily slows down, rendering it useless\textemdash.
Almost simultaneously, Huang and Wang~\cite{PhysRevB.95.035105} developed a similar scheme but using
restricted Boltzmann Machines (RBMs).
This interesting approach consists of building special types of RBMs that can use the physical properties of the
system being studied. The strategy by Huang and Wang lowers autocorrelation time
as well as raising the acceptance ratio of sampled configurations. The SLMC method has sparked interesting
new research~\cite{PhysRevB.95.241104,PhysRevB.96.161102,PhysRevB.97.205140,PhysRevB.98.041102,PhysRevE.100.043301};
recently, Nagai, Okumura and Tanaka~\cite{PhysRevB.101.115111} have shown that the SLMC is even more promising
by coupling it with Behler-Parrinello NNs~\cite{PhysRevLett.98.146401}, making it more general and
less biased towards specific physical systems.

Despite its advantages, the SLMC method still has some shortcomings and drawbacks. As with every other ML model, good and representative training data is required in order to train the model; for a complex model, i.e., if the model contains a large number of parameters to be fitted, the more data samples are needed to avoid overfiting. This might come as a problem if the original MC method used to sample configurations is already slow.
Another important drawback is that the ML model needed to approximate the probability density for the configuration set must be robust enough to handle the problem at hand, so some intervention from the user is required in order to choose a good model, but this is not simple a task because this would imply that the dyamics of the system must be known from the start; this might remove all advantages of using ML models for automated computation of the problem.
To alleviate some of these disadvantages, Bojesen~\cite{bojesen_policy-guided_2018} proposed a reinforcement learning approach, defined as the Policy-guided Monte Carlo (PGMC) method. In this methodology, the focus is now in an automatic approach to target the proposal generation for each configuration, which can be done by tuning the policy for making updates.
Showing promising results in some models, the PGMC proves to be a better candidate for automatic MC sampling, although it also has its own shortcomings, which is primarily the fact that if the available information is inadequate, the training might become unstable and the probability density will be biased, ultimately breaking ergodicity. Nevertheless, these methods prove that ML models can be used in CMP to obtain some gains in computational efficiency and simplicity making them a potentially good tool for more challenging problems.

\subsection{Outlook}
Current research is geared towards enhancing mainstream methodologies in CMP with
ML, and enabling physicists to obtain high-precision results while also providing some insight of the learning mechanisms used by the ML algorithms employed.
Another current area of research in Spin Systems is that of figuring out which
ML techniques are more suitable to understand out-of-equilibrium systems~\cite{PhysRevLett.120.257204}, or
even more complex frustrated models~\cite{PhysRevB.100.125124}, just to name a few.
Challenges, such as represetantive data set generation and feature selection must also be overcome within this area if we wish to see these methodologies become more popular. This matter is even more important for out-of-equilibrium systems because it is not a simple task to generate useful training data sets.

\section{Physics-inspired Machine Learning Theory}
\subsection{Explanation of Deep Learning Theory}
The gain in knowledge between ML and CMP is not one-sided as one might think.
ML has also benefited greatly from CMP with its thorough and rigorous theoretical frameworks that it possesses.
Despite their huge success in human-like tasks\textemdash
e.g., the classification of images~\cite{alom2018history}\textemdash, ML, and more importantly DL
are not always able to explain why this success is even possible in the first place because these models might not reach the algorithmic transparency of other classical ML models, e.g. \emph{linear models}. Schmidt \etal~\cite{schmidt2019recent} mention that representative datasets, a good knowledge of the training process, and a comprehensive validation of the model can usually overcome this obstacle. Other classical algorithms, such as MC computations are not always analytically tractable, so we might argue that this problem is not unique to ML models. With this in mind, an interpretation of the models used can be done to some extent, and even more so with the help of some theoretical frameworks, such as those found within Physics.

In 2016, Zhang \etal~\cite{zhang2016understanding}
took it upon themselves to run detailed experiments that showed the lack of complete understanding of the current
learning theory from a theoretical perspective; on the practical side of things, DL is still very prosperous in its
results. 
Since then, physicists have tried to give insights~\cite{lin2017does} into this topic, creating a fruitful
research area with extremely interesting cross-fertilization between both communities. For a more thorough and complete
analysis on the explanation and interpretability of DL contributions until now,
the reader can refer to the recent review on the topic by Fan, Xiong and Wang~\cite{fan2020interpretability}.

\subsection{Renormalization Group}
The lack of theoretical understanding of the learning theory in most ML algorithms seemed like a good challenge for
condensed matter physicists that exerted experienced manipulation in theoretical frameworks, such as the
Renormalization Group (RG)~\cite{PhysRevB.4.3174,PhysRevB.4.3184}.
RG consists of a mathematical framework that allows us a very rigorous and systematic investigation
of a many-body system by working with short-distance fluctuations and building up this information to understand long-distance
interactions of the whole system.
In 2014, Mehta and Schwab~\cite{Mehta2014} provided such an insight
by creating a mapping between the Variational Renormalization Group and the DL workflow that is common
nowadays. This turned out to be an interesting take on the problem, resulting in a very rigorous framework by which
one could explain the learning theory that Deep NNs follow when data is fed to them.
In the same spirit, Li and Wang~\cite{PhysRevLett.121.260601} proposed a new scheme, this time in a more systematic
way, by means of hierarchical transformations of the data from an abstract space\textemdash the space of features
present in the data\textemdash to a latent space. It turns out that this
approach has an exact and tractable likelihood, facilitating unbiased training with a step-by-step scheme; they
proved this through a practical example using Monte Carlo configurations sampled from the one-dimensional and
two-dimensional Ising models.
More research regarding a deep link between RG and DL has been carried out since then, for instance,
Koch-Janusz and Ringel~\cite{koch2018mutual} showed that even though the association between RG and DL
is strong, most of the practical methodology to achieve good results from this is not so readily available. By
using feed-forward NNs, they proved that a RG flow is obtainable without discriminating the important
degrees of freedom that are sometimes \emph{integrated out} by following the RG framework. This seems like a promising
and enriching area of research as new insights are constantly
developed~\cite{PhysRevX.10.011037,koch2020short,hu2019machine}.

\subsection{Restricted Boltzmann Machines} \label{sec:rbm}
Instead of giving a complete explanation of a learning theory, CMP has been able to give
thorough explanations of ML algorithms; such is the case of the \emph{Restricted Boltzmann Machine} (RBM)
~\cite{smolensky1986information}. This type of model is a very interesting case of the inspiration drawn from statistical
physics onto ML. In the RBM, a NN is trained with a given data set, the intelligent model
then learns complex internal representations of the data, after the model has been trained it outputs an approximation
of the probability distribution of the input data.
The resulting approximate probability distribution is then linked to a certain ML task,
be it \emph{classification}~\cite{10.1145/1390156.1390224}, \emph{dimensionality reduction}~\cite{Hinton504},
\emph{feature selection/engineering}~\cite{coates2011analysis}, and many more.
The underlying structure of the model is not a conventional FFNN, instead, it is a type of model that is built using 
graphs~\cite{fischer2014training}.
The principal mechanism by which the RBM learns features is by computing an \emph{energy} function $E(h,v)$
from certain \emph{hidden}\textemdash $h$ \textemdash and \emph{visible} \textemdash $v$ \textemdash units
that form a \emph{configuration}; after computing this \emph{energy} function $E(h,v)$ one evaluates the Boltzmann factor
of the energy to obtain a probability for the given configuration \textemdash $P(h,v)=e^{-E(h,v)} / Z$ \textemdash.
From classical statistical mechanics we know that the \emph{partition function}
$Z=\sum_{i} e^{-E(h,v)}$ is the normalizing constant needed to ensure that $0 \leq P(h,v) \leq 1$.
It is this link between the Boltzmann description and the energy function from the RBM that makes it an
ideal candidate to be analyzed under the statistical mechanics framework.

Interestingly enough, RBMs are extremely versatile, despite the fact that it is a ML method for which the training mechanisms
are not as fast as in more modern ML algorithms,
thus special technical schemes have been developed for the task. This is possibly the greatest drawback of using RBMs, the training procedure is not unique and the most common procedure to train such models is the so-called \emph{contrastive divergence} algorithm~\cite{hinton_training_2002}.
This algorithm can be tricky to use effectively and requires a certain amount of practical experience to decide how to tune the corresponding hyperparameters from the model.
A good practical guide to learn about training RBMs is the one by Hinton~\cite{hinton_practical_2012}, where special care is taken to inform the reader of the uses and shortcomings of the training procedure for RMBs.
We shall discuss some of the attempts at providing some insight into the training mechanisms of RMBs by using some theoretical frameworks.

For example, Huang and Toyoizumi~\cite{PhysRevE.91.050101} introduced a mean-field theory approach to perform the training
using a message-passing-based method that evaluates the partition function, together with its gradients without requiring
statistical sampling\textemdash which is one of the drawbacks of traditional learning schemes\textemdash.
Further work on understanding the thermodynamics of the model were carried out~\cite{decelle2018thermodynamics}
with encouraging results, showing that the learning mechanism in general is completely tractable, albeit complex
when the model enters a non-linear regime.
The RBM can also be seen as the inverse Ising model~\cite{doi:10.1080/00018732.2017.1341604}, where instead
of obtaining physical quantities like order parameters and critical exponents from system configurations,
one has as input the physical observables and intends to obtain insights on the system itself. With this theoretical
framework in mind, the first steps to produce rigorous explanations of the algorithm were done by
Decelle et al.~\cite{Decelle_2017} by studying the linear regime of the training dynamics using spectral analysis.

RBMs are such rich and resourceful models that have been applied to a number of hard tasks; for instance, the modeling
of protein families from sequential data~\cite{tubiana2019learning,doi:10.1162/necoa01210}. Another such complex task is that
of continuous speech recognition, on which a special type of RBM was employed with promising results~\cite{dahl2010phone}.
Within the realm of CMP, these models have been used to construct accurate ground-state wave functions of strongly interacting and
entangled quantum spins, as well as fermionic models on lattices~\cite{PhysRevB.96.205152}.

RBMs are currently one of the most flexible algorithms, despite its training mechanisms disadvantages. These models are particularly well suited to study problems within CMP and quantum many-body physics~\cite{melko2019restricted}.
With even more research efforts to make RBMs scalable to larger data set and faster training mechanims, it might be possible for RBMs to become a standard tool in ML applications to CMP problems.

\subsection{Interpretation of Machine Learning models} \label{interpretation}
Despite the success they have achieved, ML algorithms are prone to a lack of
interpretation, resulting in so-called \emph{black box} algorithms that cannot grant a full
rigorous explanation of the learning mechanisms and features that result in the outputs obtained
from them.
However, for a condensed matter physicist, it is more important to understand the core mechanism by which the
ML model obtained the desired output, so that it can later be linked to a useful physical model. As an example,
we can take the model with the Hamiltonian described by Eq.~\eref{eq:ising} and train a FFNN on simulation data, following the methodology
of Carrasquilla and Melko~\cite{Carrasquilla2017}; once the NN has been trained, the critical
exponent is obtained and used to compute the critical temperature, but the question remains, what enabled the NN
to obtain such information? Is there a mechanism to extract meaningful physical knowledge from the weights and layers
from the NN?

To answer this type of questions researchers have had to resort to arduous extraction and analysis from the
underlying model architectures. For example, Suchsland and Wessel~\cite{PhysRevB.97.174435}
trained a shallow FFNN on two-dimensional configurations from Hamiltonians similar to Eq.~\eref{eq:ising}, then proceeded to
comprehensively analyze the weight matrices obtained from the training scheme, while also performing a
meticulous study of the reasons why activation layers and their respective neurons fire a response when required;
they found out that both elements\textemdash training weights and activation layers\textemdash play a crucial
role in the understanding of how the NN is able to output the physical observables that we give a meaning to.
They also determined that the learning mechanism corresponds to a domain-specific understanding of the system, i.e.,
in the case of the Ising model, the NN effectively learned the structure of the surrounding neighbors
for a particular spin, with this information it then proceeded to compute the magnetization order parameter
as part of the computations involved in the training and predicting strategies.

Another approach is to use classical ML algorithms like SVMs given that
these models provide a rigorous mathematical formulation for the mechanism by which they learn from given
features in the data. Ponte and Melko~\cite{PhysRevB.96.205146}, as well as Gianetti et al.~\cite{Giannetti2019}
exploited these models and used the underlying \emph{decision function} of the SVMs to map
the Ising model's order parameters and obtained a robust way to extract the critical exponents from the system.
Within this approach, one does not need to provide a meticulous explanation of the learning mechanism for the model,
this has already been done in its own theoretical formulation\textemdash instead, an adjustment of the physical
model with respect to the ML algorithm needs to be performed in order to successfully achieve a link between both frameworks.
Nonetheless, this approach has the disadvantage that it can be hard to map the ML
method to the physical model, as it requires a deep and experienced understanding of the physical system, while
simultaneously needing a comprehensive awareness of the theoretical formulation of the ML algorithm
being used.

\subsection{Outlook}
The approach of illustrating the training scheme, as well as the loss function landscape
of a ML algorithm with CMP has proven to be both intriguing and effective.
Baity-Jesi et al.~\cite{baity2019comparing} used the theory of
glassy dynamics to give an interesting interpretation of the training dynamics in DL models.
Similar work was performed by Geiger et al.~\cite{PhysRevE.100.012115}, but this time with the theory of
jamming transitions, disclosing why poor minima of the loss function in a FFNN
cannot be encountered in an overparametrized regime\textemdash
i.e., when there exists more parameters to fit in a NN than training samples\textemdash.
Dynamical explanations of training schemes have the advantage of providing precise frameworks that
have been understood in the physics community for a long time, but it is an intricate solution as it needs
a scrupulous analysis of the model, leaving little room for generalization on other ML models.
It is true that FFNNs, as well as Deep NNs are the less understood ML
models, but they are not the only ones that need a carefully detailed framework; maybe in future research
we can see how other ML models benefit from all these powerful techniques developed so far by
the CMP community.

\section{Materials Modeling}
\subsection{Enhanced Simulation Methods}
Molecular dynamics (MD) is a classical simulation technique that integrates Newton’s equations of motion
at the microscale to simulate the dynamical evolution of atoms and molecules~\cite{allen2017computer}.
It is, along with the MC method, a fundamental tool in computer simulations
within CMP, specifically in HM, SM and Quantum Chemistry. The powerful method of MD enables
the scientist to obtain physical observables through a controlled and simulated environment, by-passing
limitations that most experimental setups have to deal with. MD has become a cornerstone method to model complex
structures, such as proteins~\cite{LINDAHL2008425} and chemical structures~\cite{doi:10.1021/ar9702841}, just
to name a few. But such complex structures have to deal with large number of particles\textemdash with their
many degrees of freedom\textemdash as well as complicated interactions in order to obtain meaningful results
out of the simulations. This creates a difficult challenge to overcome, as efficient exploration of the phase space is a difficult task, even with MD simulation code accelerated with modern hardware. In order
to reduce the computational burden of modeling such many-body systems, special enhanced sampling methods
have been developed~\cite{doi:10.1063/1.5008853}, but not even these schemes can alleviate the arduousness
of modeling complicated systems. ML is seen as a promising candidate to surpass modern solutions
in MD simulations, as shown primarily by Sidky and Whitmer~\cite{doi:10.1063/1.5018708}. In their novel
approach, authors employed FFNNs to learn the free energies of relevant chemical
processes, effectively enhancing MD simulations with ML and creating a comprehensive method
to obtain high-precision information from simulations that were not previously readily available.
Another interesting procedure to enhance MD simulations was the automated discovery of features that could be
potentially fed into intelligent models, making it possible to create a systematic and automated workflow
for materials research. Sultan and Pande~\cite{doi:10.1063/1.5029972} developed such techniques by training several
classical ML methods \textemdash Support Vector Machines, Logistic Regression, and
FFNNs \textemdash, showing that given a MD framework one could set up a complete,
self-adjustable scheme to obtain the system's free energy, entropy, structural information and many more observables.

Another powerful tool to study many-body systems is Density Functional Theory~\cite{marx2009ab} (DFT), which is a quantum mechanical framework for the 
investigation of the electronic structure of atoms. We refer the reader to useful 
references for more information on DFT~\cite{koch2015chemist,parr1980density}.
Although a very robust tool its main disadvantage is that the performance of DFT results depends on the approximations and functionals used; it is also computationally demanding for large systems.
The use of NN to enhance these computations has shown favorable results, for instance, in the work by Cust{\'o}dio, Filletti and Fran{\c{c}}a~\cite{custodio2019artificial}, where an approximation of a density functional by means of a NN reduces the computational cost, while simultaneously obtaining accurate results for a vast range of fermionic systems.
For similar applications to DFT, we would like to refer the reader to a review where a number of different methods are used to solve the Schr{\"o}dinger equation and related problems of constructing functionals for DFT~\cite{Manzhos_2020}.

Although in its early stages, the enhancement of classical methods with ML is an intriguing area of
research. When no rigorous foundation is needed, ML models are able to exceed human performance in
the modeling of atomistic simulations, making it a suitable candidate to become the mainstream method in the future.
Of course, there is a long path ahead to fully embrace these methods, as a complete simulation strategy is not
easily available yet; further research on the coupling between molecular simulations and ML schemes
is the primary way to achieve this.

\subsection{Machine-learned potentials}
In this new day and age, computer simulations are a standard technique to do research in CMP
and its sub-fields. These simulations produce enormous quantities of data depending on the
sub-field they are tasked in, for example, biophysics and applied soft matter have produced one of the
largest databases\textemdash the Protein Data Bank~\cite{wwpdb2019protein}\textemdash that contain 
free theoretical and experimental molecular information as a by-product of research in specific topics. This
explosion of immense data available to the public has attracted scientists to look into the ML
methodologies that have been created for the sole purpose of dealing with the so-called
\emph{big data outbreak}~\cite{Hilbert60} of recent years, and effectively uniting both fields to attempt to
use such well-tested intelligent algorithms in the understanding and automation of computer-aided models.

One such task is that of constructing complex \emph{potential energy surfaces} (PESs) with the experimental
and computational data available~\cite{Behler2016}. This has turned out to be a successful application to
materials science, physical chemistry and computational chemistry, as it allows scientists to obtain high-quality,
high-precision representations of atomistic interactions of all kinds of materials and compounds. In atomistic
simulations and experiments, scientists are always interested in obtaining a good representation of PESs because
it guarantees the full recovery of the potential energy\textemdash and as a consequence, the full atomistic
interactions\textemdash of a many-body system. The baseline technique that provides such information is the coupling
of MD with DFT which turns out to be a very computational demanding scheme, even for the simplest cases. This constituted the starting point in a
long standing approach to use ML models that could compute such complicated functions, but not without the usual challenges of good data representation, stable training and good hyperparameter tuning.
The first attempt of this was done by Bank \etal~\cite{doi:10.1063/1.469597} where they proposed to fit
FFNNs with information produced by some low-dimensional models of a CO molecule chemisorbed on a Ni(111) surface.
They obtained promising results, such as faster evaluation of the potential energy by means of the
NN model compared to the original empirical model. The data sets used by Bank \etal were composed with
a range from two up to twelve features, or degrees of freedom, making it easy and simple to train the NN models.

But this is not always the case, in fact, one could argue that it is \emph{never} the case, as systems like
proteins consist of thousands of particles and electronic densities, hence a large number of degrees of
freedom need to be computed simultaneously~\cite{ozboyaci_kokh_corni_wade_2016}.
Another challenge for these type of ML potentials is that of the \emph{descriptors} or \emph{features}
needed in order to be fed to an intelligent model. Data fed into ML models should account for all possible variations
and transformations seen in real physical models to extend the generalization of the ML schemes and avoid
bias in them.

It took more than ten years after the pioneering work of Bank \etal for Behler and
Parrinello~\cite{PhysRevLett.98.146401} to undertake this challenge to a new level
by proposing a similar methodology employing FFNNs, advocating now to a more generalized
approach to computing the PES of an atomistic system by creating what they called \emph{atomic-centered symmetric functions}. We will describe in detail this method as it is an important achievement in the application of ML to CMP and materials modeling.

The main purpose of the Behler-Parrinello approach is to represent the total energy $E$ of the system as a sum of atomic contributions $E_i$, such that $E = \sum_{i} E_i$. This is done by approximating $E$ with
a NN. The input of the NN should reflect information of the system from the contribution of all corresponding atoms, while also preserving important information about the local energetic environment.
In this sense, symmetric functions are used as \emph{features}, transforming the positions to other values that contain more useful information from the system.
In order to define the energetically relevant local environment, a cutoff function is employed $f_c$ of the interatomic distance $R_{ij}$, which has the form

\begin{equation}
    \label{eq:cutoff}
    f_c (R_{ij}) =
    \cases{
        0.5 \cdot \left[ \cos{\left( \case{\pi R_{ij}}{R_c} + 1 \right)} \right] & for $R_{ij} \leq R_c$ , \\
        0 & for $R_{ij} > R_c$ . \\
    }
\end{equation}

Next, radial symmetry functions are constructed as a sum of Gaussian functions with parameters $\eta$ and $R_s$

\begin{equation}
    \label{eq:radial}
    G_{i}^{1} = \sum_{j \neq i}^{all} \e^{-\eta (R_{ij} - R_{s})^2} f_c (R_{ij}) .
\end{equation}

Finally, angular terms are constructed for all triplets of atoms by summing the cosine values of the angles $\theta_{ijk} = \case{\bi{R_{ij}} \cdot \bi{R_{ik}}}{R_{ij} R_{ik}}$ centered at atom $i$, with $\bi{R_{ij}} = \bi{R_{i}} - \bi{R_{j}}$,
and the following symmetry function is constructed

\begin{equation}
    G_{i}^{2} = 2^{1 - \zeta} \sum_{j,k \neq i}^{all} \left( 1 + \lambda \cos{\theta_{ijk}} \right)^{\zeta} \e^{-\eta (R_{ij}^{2} + R_{ik}^{2} + R_{jk}^{2})} f_c (R_{ij}) f_c (R_{ik}) f_c (R_{jk}) ,
    \label{eq:angular}
\end{equation}

with parameters $\lambda (= +1, -1)$, $\eta$, $\zeta$. These parameters, along with the ones defined in Eq.~\eref{eq:radial} need to be found before these symmetry functions can be used as input to the NN. The parameter fitting is done by doing DFT calculations for the system to be studied.
By computing the root mean squared error (RMSE) between the fit and the true values from the computations, the training set is constructed with the DFT calculations if the RMSE is larger than the fit.
With this information, a training and testing data set is constructed and then used to train the NN, which in turn will yield an approximation for $E$.
This approximation is then used to compute different observables and quantities from the system.

This methodology introduced a paradigm shift because it was no longer a proof-of-concept that ML algorithms
could be used in these research fields, and many other attempts have been carried out since then. Following the
footsteps of Behler and Parrinello, Bart\'ok \etal~\cite{PhysRevLett.104.136403} proposed a similar methodology, but
with completely different ML schemes\textemdash in this case being Gaussian Processes~\cite{williams2006gaussian}
\textemdash which in turn needed new atomistic descriptors. We shall discuss atomistic descriptors in more detail in
the next section.

The advantages of such techniques in all the sub-fields of CMP is manifold, the main one
being that simulations are now faster and more precise. Another advantage of such methods is the fact that we can leverage the
retention of the learned features from the NN and by employing \emph{transfer learning}~\cite{5288526}
the model does not need to be trained again to compute the same interactions it learned to compute. For the reader interested in more detailed discussions on transfer learning we point out some reviews~\cite{torrey_transfer_2010,weiss_survey_2016} that give enough information on the current advancements on this topic.
Nevertheless, this approach has one significant drawback which is the fact that most NNs cannot learn sequential tasks, which in turn creates a phenomenon called \emph{catastrophic forgetting}~\cite{french_catastrophic_1999,mccloskey_catastrophic_1989,kumaran_what_2016,mcclelland_why_1995}. In the case of catastrophic learning, the NN abruptly loses most of the information learned from a previous task, unable to use it in the new task at hand.
Transfer learning is a ML technique that needs meticulous manipulation from the user, involving \emph{fine tuning}, which is a training process where some of the learned information are kept and some are learned again, and then \emph{joint training} where new information is now learned using previous and newly acquired information obtained from fine tuning.
Research is currently being carried out~\cite{kirkpatrick_overcoming_2017,li_learning_2018} to overcome this disadvantage from NNs and shared learning.

Machine-learned potentials have also been successfully applied to explaining complex interactions, like the van der Waals
interaction in water molecules and the reason this interaction results in such unique properties
seen in water~\cite{Morawietz8368}. Another interesting application was the simulation of solid-liquid
interfaces with MD~\cite{C6CP05711J}, an application that was not thought of being computationally
simple or even tractable. One of the remaining challenges in this area of research is a paradoxical one, being that
these ML potentials require very large data sets that are computationally demanding in order to train
the intelligent model, one possible approach to alleviate this issue
is to build and maintain a decentralized data bank similar to the Protein Data Bank\textemdash which
could be a cornerstone of modern computational materials science and chemistry\textemdash enabling anyone interested in this
type of research to collect insights from the acquired data, without needing to run large-scale simulations
or re-train complicated NN architectures to do so.
This is an interesting research field in the junction of many other specialized sub-fields which
is also growing faster due to current computing processing power available, setting up the pathway for
new and interesting intelligent materials modeling\textemdash which could surely become the future of materials
science.

\subsection{Atomistic Feature Engineering}
So far, we have discussed the different ways in which scientists have been able to apply ML
technology in different sub-fields of CMP, but lattice systems seem to prevail, where most
of the research done is reported with these systems\textemdash why is that?\textemdash.

One of the fundamental parts of using a ML method is that of \emph{data input} with specific
features; for instance, images are described by \emph{pixels} in a two-dimensional grid where the pixels
are the features that numerically represent the image, these pixels are then fed to a ML model, e.g., a classifier;
another clear example are numerical inputs that can be readily used in regression tasks, and then used as
a prediction model. We wish to obtain a similar representation for physical systems, so as to
ease the use of ML models in condensed matter systems.
Recall that the intrinsic features of lattice systems are spins ($\sigma_i$), these spins do not change location
they actually stay put and only change their spin value taken from a finite, discrete set of values.
We can easily see a link between images and pixels, and a lattice system and its spins;
both can be described by a set of discrete features that do not change their values as the system is evolving and changing.
But we cannot say the same about atomistic and off-lattice systems, thus we cannot use the same representation
as lattice systems because
atomistic systems are described primarily by the constant movement of their atoms and particles, and their many
degrees of freedom\textemdash for instance,
for a three-dimensional system, one has $3N$ translational degrees of freedom with $N$ being the total number of particles\textemdash.
By using the raw positional coordinates, velocities or accelerations, we might not be able to convey the true features
that the system possesses; if even possible, they could be conveyed in a \emph{biased} way, e.g., by defining an arbitrary
coordinate system or by not being invariant to translation and rotations.
In order to overcome this limitation, a special mechanism needs to be developed to feed this data to the ML algorithms. It is this \emph{encoding} of raw atomistic data
into a set of special features that has seen a strong surge in recent years when dealing with materials of any kind.

Atomistic descriptors have been studied heavily throughout the years~\cite{PhysRevB.87.184115,PhysRevLett.108.058301,THOMPSON2015316,doi:10.1063/1.4825111,C1CP00051A,PhysRevB.89.205118,doi:10.1002/qua.24917}
because they are the gateway to applying ML methodologies in atomistic systems. The reader is referred
to these reviews~\cite{Behler2016,ceriotti2019} for more information on the topic. Descriptors are a systematic way
of encoding physical information in a set of \emph{feature vectors} that can then be fed to a ML algorithm.
These encodings need to have the following properties: a) the descriptors must be invariant to any type of point symmetry,
for example, translations, rotations and permutations; b) the computation of the descriptor must be \emph{fast} as it needs
to be evaluated for every possible structure in the system; and c) the descriptors have to be differentiable with respect
to the atomic positions to enable the computation of analytical gradients for the forces. Computing these descriptors
is essentially called \emph{feature selection} in the ML literature, and most of the time these procedures
are carried out in an automated way in a normal ML workflow. Once a particular descriptor has been used,
all the methodology from ML that we have seen applied to lattice systems can now be employed on atomistic
and off-lattice systems.

\subsection{Outlook}
ML-enhanced materials modeling is a strong research area, with a focus on streamlining computations and automating material discoveries. Descriptors that employ less data should be the primary research area, as it currently is one of its main shortcomings.
A standard framework and benchmark data sets could also be used to further this area, while also looking forward to an automated workflow.

\section{Soft Matter}
At this stage, most of the ML applications we have discussed have primarily been focused on HM and atomistic many-body systems.
It has been so due to the fact that ML methods can be applied using data sets that need little feature engineering, even without its disadvantages.
However, off-lattice and SM systems, along with other atomistic many-body systems, on the other hand, need thorough feature engineering and a different approach to training ML models.

The applications we will review in this section reflect this inherent challenge of finding representative features that could enable ML models to be used in tasks common to all CMP, e.g., phase transitions and critical phenomena.

\subsection{Phase Transitions}
Phase transitions are the cornerstone of ML applications to lattice systems, but with the use of atomistic descriptors these can easily be extended to off-lattice systems, which are usually observed in SM. 
Jadrich, Lindquist and Truskett~\cite{jadrich2018unsupervised1,jadrich2018unsupervised2} developed the idea of using the set of distances between particles as a descriptor for the system, and used unsupervised learning\textemdash in particular, Principal Component Analysis\textemdash on these descriptors.
They wanted to test whether a ML model was able to detect a phase transition in hard-disk and hard-sphere systems.
When the unsupervised method was applied, the output from the model was later found to be the positional bond order parameter for the system~\cite{PhysRevE.73.065104,PhysRevLett.76.255}, which effectively determines whether the system has been subjected to a phase transition.
In other words, the ML model was able to automatically compute the order parameter without having been told to do so.
A similar approach to colloidal systems was done by Boattini and coworkers~\cite{doi:10.1063/1.5118867}.
Instead of using inter-particle distances, they computed the orientational bond order
parameters~\cite{PhysRevB.28.784,doi:10.1063/1.2977970}, then they used a DL unsupervised method\textemdash a FFNN based autoencoder\textemdash to obtain a latent representation of the computed order parameters.
With all these preprocessing steps, they fed the encoded data to a clustering algorithm and obtained detailed descriptions of all the phases of matter in the current system. 
This methodology is systematic and leads itself nicely to an automated form of phase classification in colloidal systems.

\subsection{Colloidal systems}
Liquid crystals are a state of matter that have both the properties of a liquid and of a solid crystal~\cite{chandrasekhar1992liquid}. At the simplest case, liquid crystals can be studied using one of the reference model from SM, i.e., hard rods, or similar anisotropic particles~\cite{sluckin2004crystals}.
This type of systems exhibit properties similar to those present in HM, such as topological defects.
The main issue with using the same methodology is that building a data set with useful and representative features in systems made of liquid crystal molecules is not trivial, mainly due to the fact that positions and other physical observables can take any continuous value instead of a specific value from a given set.
In the work by Walters, Wei and Chen~\cite{walters_machine_2019} a time series approach using Recurrent NN was employed, using the basic descriptors from a system of liquid crystals, namely the orientation angle and positions of all the particles in the system.
By using long short-term memory networks~\cite{lstm1997,lstm2020} (LSTM), they showed that the off-lattice simulation data can be used as input and expect to identify between several topological defects.
This is a great step towards the use of mainstream techniques from ML in SM, as one of the main drawbacks, which is feature selection and data set building, is resolved by using this kind of methods. Although a data set collected from computer simulations does not convey much problem, it might be so for experimental setups.

Similarly, Terao~\cite{doi:10.1080/1539445X.2020.1715433} used CNNs to analyze liquid crystals, as well as different types of solid-like structures, by computing special descriptors for the system.
In particular, the approach presented in that work can be seen as a generalization of the one tipically used for HM, which is to transform the off-lattice data into a common structure normally used in computer vision.
By extracting any subsystems with few particles from the total system, a conversion to grayscale images of size $l \times l$ in the case of 2 dimensional systems were performed by using the following expression,

\begin{equation*}
    u_{i,j} = C \sum_{n=1}^{m_p} \mathrm{exp} \left(- \frac{\vert \mathbf{r}_n - \mathbf{R}_{ij} \vert^2}{a^2} \right) ,
\end{equation*}

were $\mathbf{r}_n$ is the positions of the $n$-th particle in the subsystem with a size $L_s$ and total number of particles $m_p$; $\mathbf{R}_{ij}$ is the positional vector in the center of the pixel $(i, j)$; and the constant $C$ is determined such that the set $\{u_{i,j}\}$ is rescaled within the interval $[0,1]$.
The advantage of this method is its generalization potential due to the fact that no physical input is needed in order for the technique to work, so a large range of systems could be studied under the same methodology.
One possible disadvantage is the fact that some systems might need to conserve some kind of symmetries, so this should be taken into account when constructing the data set.

Polymers are materials that consist of large molecules composed of smaller subunits. These materials are of great interest in SM because most polymers are surrounded by a continuum, turning them into colloid systems~\cite{de1979scaling}.
Polymers are also seen in nature, e.g., the case of the deoxyribonucleic acid (DNA) found in almost all living organisms.
Even though these systems have been studied extensively throughout the years, computational and experimental issues always arise. It is in these situations where ML might be a suitable approach.
One such example is the design of a computational method for understanding and optimizing the properties of complex physical systems by Menon \etal~\cite{menon_elucidating_2017}.
In the work by Menon \etal physical and statistical modeling are integrated into a hierarchical framework of machine learning, as well as using simple experiments to probe the underlying forces that combine to determine changes in slurry rheology due to dispersants.
One of the main contributions of this work is that they show how physical and chemical information can aid in the reduction of the number of samples in a data set needed to do efficient ML modeling.
If data sets are constructed using experimental data and some insight about the problem, the use of ML techniques might be easier to achieve.
An issue presented in the work by Menon \etal considers how data sets might be subjected to correlation between system variables and responses from the system. This could be a potential issue when training some ML models due to training stability and yielding acceptable results.

On a more recent approach to polymers one can find is the work by Xu and Jiang~\cite{xu_machine_2020}, where a ML methodology is used for studying the swelling of polymers immersed in liquids.
In this work, the importance of using chemical and physical descriptors from the beginning to construct the ML framework is highlighted.
By constructing and using these descriptors, molecular representations are used, along with data augmentation, to study the relationships between molecular fragments and swelling degrees with the help of PCA.
Various other applications are constantly arising, such as the discovery and design of new types of polymers in ML methods~\cite{wu_machine-learning-assisted_2019}. As ML methods help with complex representations of data, it could potentially be used to accelerate the discovery of innovative materials within polymer science.

\subsection{Properties of equilibrium and non-equilibrium systems}
The structural and dynamical properties of equilibrium and non-equilibrium systems, such as complex liquids, glasses, gels, granular materials, and many more, are extensively studied under the branch of colloidal SM~\cite{hamley2013introduction,jones2002soft}.
Common tools, such as computer simulations are used to study such properties, but in some systems these tools are computationally demanding to the point that special facilities, such as high-performance computing (HPC) clusters are needed.
Even if in the current era we are able to use this kind of computational resources, we are still limited to the amount of computations needed for large, physically-representative systems.
Studying simple reference models, Moradzadeh and Aluru~\cite{moradzadeh_molecular_2019} were able to fabricate a DL autoencoder framework to compute structural properties of these simple systems.
Using MD and long running simulations, they collected several position snapshots for a diverse range of Lennard-Jones systems and using the autoencoder as a denoising framework they computed the time average radial distribution function (RDF).
This was achieved using a data set built with three main features, the thermodynamic state composed of the temperature and density\textemdash $(T, \rho)$ \textemdash at each time step, and the RDF snapshot at that exact moment.
Both time and data are efficiently reduced with this methodology, as only 100 samples are needed when the network is trained, although a large data set is employed to do such training.

Non-equilibrium systems have also seen some applications from ML. Employing SVMs, Schoenholz \etal~\cite{schoenholz_structural_2016} found that the structure is important to glassy dynamics in three dimensions.
The results were obtained by performing MD simulations of a bidisperse Kob–Andersen Lennard-Jones glass~\cite{PhysRevLett.73.1376} in three dimensions at different densities $\rho$ and temperatures $T$ above its dynamical glass transition temperature.
At each density, a training set of 6,000 particles was selected, taken from a MD trajectory at the lowest $T$ studied, to build a hyperplane in $\Re^{M}$ using SVMs.
This hyperplane is then used to calculate $S_i (t)$ for each particle $i$ at each time $t$ during an interval of 30,000 time steps at each $\rho$ and $T$, where $S_i$ is the \emph{softness} of a particle $i$, defined as the shortest distance between its position in $\Re^{M}$ and the hyperplane, where $S_i > 0$ if $i$ lies on the soft side of the hyperplane and $S_i < 0$ otherwise.
With this method, it is shown that there is structure hidden in the disorder of glassy liquids and this structure can be quantified by softness, computed through a ML framework.

New ML applications to design and create glass materials are steadily appearing as presented in the review by Han \etal~\cite{liu_machine_2019}, where challenges and limitations are also discussed for these applications to glassy systems.

\subsection{Experimental setups and Machine Learning}
Experiments are a fundamental part of CMP, providing results which can be later compared to theoretical and computational results. Experimental setups have also been enhanced with ML models in different aspects of SM.

In the work by Li \etal~\cite{li_machine_2018} collective dynamics are explored, measured through piezoelectric relaxation studies, to identify the onset of a structural
phase transition in nanometer-scale volumes. Aided by $k$-means clustering on raw experimental data, Li \etal were able to automatically determine phase diagrams on systems where identification of underpinning physical mechanisms is difficult to achieve and microscopic degrees of freedom are generally unavailable for observations.
Although these results are more precise than those obtained with traditional tools, experimental setups might still be difficult and expensive to perform each time, so ML methods might be able to alleviate this shortcommings by requiring less data or performing more efficiently than other techniques.

We have touched several times on the fact that data and the amount of samples needed are one of the main shortcomings for the applications of ML methods to CMP. This is particularly more of a challenge on experimental setups, where large quantities of data samples might not be possible on a given system.
The end-to-end framework developed by Minor \etal~\cite{minor_end-end_2020} shows that this problem can be overcome. The collection of physical data of a liquid crystalline mesophase was performed using an experimental setup, i.e., a mechanical quench experiment.
The ML pipeline consisted of real time object detection using the YOLOv2~\cite{redmon2016yolo9000} architecture to detect topological defects on the system. Standardization and image enhancement techniques were needed to mimic real-world inaccuracies in the training data set.
The model resulted in comparable spatial and number resolution human annotations to which the pipeline was compared; with significant improvement on the time spent in each frame, thus resulting in a dramatic increase in the time-resolution (more frames analyzed).

\subsection{Outlook}
As we have discussed so far, the fundamental step to applying ML to SM and off-lattice systems
is to define and use a set of meaningful descriptors, but the choices for descriptors are large as well as system-dependent. In the case of experimental setups, the efficient use of data samples are necessary.
In order to fully embrace ML in SM and colloidal systems a more automated approach will need
to be developed, posing it as new and fascinating research area. Recently, DeFever \etal~\cite{C9SC02097G} developed
a generalized approach that does not use hand-made descriptors; instead, a complete DL workflow
is used with raw coordinates from simulations and it is able to obtain promising results about the
phase transitions, both in and out of thermodynamic equilibrium in SM systems.
This scheme is a great step forward towards a fully automated strategy
for off-lattice systems, as it reduces the amount of bias a scientist might introduce while using certain
descriptors for given systems. 
The only downside to employing these types of DL methodologies is the
large amount of data needed, so maybe in future research we can see approaches where one can use a small amount of data
to obtain similar results. These disadvantes seem to disappear in end-to-end frameworks as the one described in Minor \etal~\cite{minor_end-end_2020}; still, the amount of data might be an obstacle when applying ML models. This is discussed in detail in the next section.

\section{Challenges and Perspectives}
At this point of the review, we have discussed some of the major applications of ML to CMP. We have also touched on other useful insights provided by theoretical frameworks on ML, as well as discussing important points on various shortcomings and drawbacks of using ML models in diverse applications.
In this section, however, we would like to expand on such discussions and address the main challenges that we see in the current climate of this research area.

\subsection{Benchmark data sets and performance metrics}
In order to test new algorithms in ML a \emph{benchmark data set} is needed. This data set serves the purpose of assessing the performance of a particular algorithm by means of a defined \emph{performance metric}. One such example is the extremely popular MNIST data set~\cite{lecun_gradient-based_1998}, which is currently extensively used to test new ML algorithms.
The most common task to solve on this data set is to check if the proposed ML algorithm can successfuly recognize all ten different classes of handwritten digits, in a supervised learning fashion. Thus, the performance metric is the \emph{accuracy}\textemdash which corresponds to the total number of correct predictions over the total number of samples in the data set. This means that if a proposed NN architecture scores a $97 \%$ accuracy on the MNIST data set, the NN architecture can predict at least $9.7$ handwritten digits correctly. Although the MNIST has been widely used, it is not the only one and most researchers agree that this data set should be replaced by a more challenging one, for instance the Fashion-MNIST~\cite{xiao_fashion-mnist_2017}.

With this in mind, the main point we wish to convey is that there is an important need for standardized and well-tested benchmark data sets in the current research area of ML and CMP, along with their respective performance metrics.
Benchmark data sets will be used to try and test new ML models, while verifying if such models can provide the expected results, and achieve a good score on the respective performance metric.
We can look at the example of computing the critical temperature of an Ising model where we do not have the external field shown in Eq.~\eref{eq:ising}.
In this example, we have seen several possible ML models\textemdash NNs, SVMs, DL approaches, Reinforcement Learning approaches\textemdash that can achieve good results on this problem, so we might agree that such problem can be taken as a benchmark problem.
The Ising model is a simple one, which can be computed with standard MC methods and a standard benchmark data set is fairly easy to obtain.
This is not the case, however, of other CMP systems and problems, for example, the case of machine-learned potentials.
As discussed in Ref.~\cite{Behler2016}, the construction of such models currently require very large reference data sets from electronic structure calculations, making the construction computationally very demanding and time consuming.
This means that it is not simple nor easy to test newly created ML models that can tackle this problem. Thus, it implies that there is a strong need to spend countless computing hours to obtain representative data samples just to test if the proposed ML method can perform well at the simplest cases. Other CMP systems are even more complicated to produce data sets, in particular those that come from experimental setups. Experiments can be very costly; obtaining clean, labeled and representative data is not something that is currently being addressed completely.

The solution to this problem is to generate well-tested standard data sets for each problem in CMP. Although this is not a simple task, some efforts have been done in current research areas. Related to materials modeling and CMP, MoleculeNet by Wu \etal~\cite{wu_moleculenet_2018} is a contribution that curates multiple public datasets, establishes metrics for evaluation, and offers high quality open-source implementations of multiple previously proposed molecular featurization and learning algorithms.
Other effort, although in the realm of quantumm chemistry, is the QM9 data set~\cite{ramakrishnan_quantum_2014} for molecular structures and their properties.
Not much attempts are seen in the realm of SM, where one could test for the simplest cases, namely, the hard-sphere model, binary hard-sphere mixtures and some other simple liquids benchmarks.
This could signal a different approach to solving the problem of performance metrics and benchmark data sets by means of physical correctness and standard results drawn from physical theories. For instance, when testing new molecular dynamics algorithms, the Lennard-Jones potential~\cite{verlet_computer_1967} is a well established benchmark given that all its properties are well known and studied. 
Instead of having to compute these values each time, one wishes to test a new ML model, or having to read and extract the data set from the literature, a joint effort of building a common and curated data bank might help scientists to quickly test their newly proposed algorithms.

A final approach to solving this problem could be the reuse and sharing of created data sets for a given problem. While solving some problem within CMP, the data set created can be shared through an open service, e.g., Zenodo~\cite{noauthor_zenodo_nodate}, where other scientists can easily access such information and use them in an \emph{open science} way~\cite{allen_open_2019}.
Benchmarks are most needed for the prevalence and proliferation of newly proposed ML algorithms applied to CMP, and this key challenge should be one of the main challenges to overcome.

\subsection{Feature selection and Training of ML models}
Another key challenge in the application of ML to CMP is that of feature selection. ML models rely on a data set that is representative of the system itself, i.e., without interpretation capabilities, models might not provide physical insights or meaning.
If a data set is not correctly labeled or structured, training can be unstable and different issues can arise. In plain ML various feature selection techniques are available along with some useful metrics~\cite{cai_feature_2018}.
Feature selection strongly depends on the data set and problem alike, so most of the feature selection methods cannot be transfered directly to CMP problems. For this to be employed, physics criteria and insight into the problem is mandatory. For example, it requires the expertise of a SM physicist to identify betwen a nematic phase and a smectic phase in a crystal liquid, or the distinction of symmetries in a given system that need to be conserved when using a ML model.
In this review, we have seen some of these shortcomings, when talking about atomistic descriptors and other applications. Unfortunately, this is a challenge that is not so easily resolved as it requires having experience on the problem itself and the ML model to be used.
One such example that we have already covered is the description of hard-spheres and hard-disks using the distance between particles shown in Ref.~\cite{jadrich2018unsupervised1}. 
This feature selection is not a standard one in ML, it comes directly from the Physics of the problem at hand, and is an important contribution to feature selection in ML and CMP.
Research of this kind might be needed even more than creating new ML models, so that more CMP systems can be tested with more feature selection techniques. Another possible approach to such problem can be the automatization of feature selection by means of ML models; for instance, using a modification of the one proposed by Bojensen~\cite{bojesen_policy-guided_2018}.
It is interesting to see that most automatization methodologies use Reinforcement Learning, which might indicate that it is well suited to solve some of these drawbacks.

We now discuss the issue of training and using ML models in general. Handling and training ML models well so as to obtain good results is not so simple at first. Nowadays, there are a lot of standard techniques used to train ML models, such as \emph{regularization}, \emph{early stopping} and many others; for a thorough and detailed explanation of these techniques we refer the reader to the excellent review by Mehta \etal~\cite{mehta_high-bias_2019} especially written for physicists.
The main point here is that training ML models has its difficulties and most of the time is not so easy to implement all of them as not all are needed in order for the training to be stable enough.
Dealing with overfitting and underfitting is one of the main issues in training ML models, especially in CMP applications because of the possible lack of data samples and the difficulty of choosing the correct ML algorithm.
Moreover, tuning of hyperparameters in ML models requires time and patience from the user as it can be a time consuming task, especially if approaching the problem with a ML model for which there is little knowledge of how to use it correctly. There are solutions to this problem, such as Bayesian optimization~\cite{snoek_practical_2012}, a robust tool which can help tune hyperparameters in an automated fashion, with good results. Fortunately, more and more research is being carried out to alleviate the problem in ML, for instance, the DropConnect approach~\cite{wan_regularization_2013}. which greatly helps with the overfitting problem.

In general, most ML models are not so readily applicable to CMP problems, mostly because of the lack of good data representation and the need of model tuning.
There is a diverse set of possible solutions to these problems brought directly from ML, but these need to be tested in CMP problems to ensure such techniques are viable. New research should be carried out in this way to enhance current and recently proposed ML applications to CMP, and to make it easier for physicists to employ ML models.

\subsection{Replicability of results}
It is a known fact that current ML research faces a reproducibility crisis~\cite{hutson_artificial_2018}\textemdash the fact that even with detailed descriptions from the original paper, an independent implementation of such models does not yield the same results as those claimed in the original work.
One of the main problems with this phenomenon is the fact that most research papers do not provide the code that was used to obtain the results presented in the work.
Even if the code is provided, most of the results claimed cannot obtained due to the fact of stochasticity of certain algorithms\textemdash for example, when using \emph{dropout}~\cite{srivastava_dropout_2014}, a technique that reduces overfitting by randomly turning on or off certain parts of the NN architecture.
The intrinsic randomness of most ML models is another key issue in solving the replicability problem. Most researchers deal with this problem by sharing the trained models along with the code used. This can greatly alleviate the problem, the fact is that this is a rare practice in current ML research. Nonetheless, important efforts are carried out to solve this problem. OpenML~\cite{OpenML2013} is an iniciative to collect and centralize ML models, data sets, pipelines and most importantly reproducible results.
Another great example is DeepDIVA~\cite{alberti_deepdiva_2018}, an infrastructure designed to enable quick and intuitive setup of reproducible experiments with a large range of useful analysis functionality.
Current research in this area~\cite{forde_scientific_2019,10.1371/journal.pbio.3000246,samuel_machine_2020} is necessary to create outreach on research and encourage practices that can lead to reproducible science.

Code and data sharing is not a common practice within CMP. Some researchers believe that the original paper must suffice to reproduce the results claimed, this might be true for most of the proposed methodologies and techniques, but no so much in ML applications to CMP.
In order for ML to work with CMP, reproducibility is a key factor that needs to be present from the beginning. By focusing on sharing the full training pipeline, the data set used to train a given ML model, and the code used, ML can be quite succesful in CMP, most importantly, it can be implemented by anyone that wishes to use such tools.
Until now, there is no approach to this in the current climate of ML and CMP, so it is a great opportunity to start implementing these practices of code, data and model sharing to ensure reproducibility of results.

\subsection{Perspectives}
ML and CMP have created a very enriching and prosperous joint research area.
Modern ML techniques have been applied to CMP with outstanding results, and Physics has given
fruitful insights into the theoretical background of ML. But in order to embrace and fully adopt ML tools into standard
physics workflows, some challenges need to be overcome.

Currently, there is no standard way to couple both ML and Physics simulation software suites;
the main approach is to use available software and hand-craft code that can use the strengths of both
parts. Physicists have seen the great advantage that is the use of standard simulation software suites
like LAMMPS~\cite{plimpton1993fast} or ESPResSO~\cite{weik2019espresso}, but these are not meant to be used
with ML workflows. A universal, coupling scheme might be able to simplify research in this
exciting scientific area.

Quantum systems have also seen a great deal of advances with ML applications~\cite{carrasquilla2020machine},
thus it might be interesting to see if the same schemes can be modified and applied to larger-scale systems,
such as soft materials, as we have seen it was possible in the topic of phase transitions.

New interesting theoretical insights have also been proposed to explain ML by applying not only
CMP, but other Physics frameworks, like the $\mathrm{AdS}/\mathrm{CFT}$
correspondence~\cite{PhysRevD.98.046019}. Physics is full of these rich and rigorous frameworks and new research
into the area might be able to give a full and thorough explanation of the theory of Deep Learning and its
learning mechanisms.

All in all, ML and CMP are just starting to convey ideas to each other,
creating a fertile research area where the work done is towards an automated, faster and simplified workflow that will
enable condensed matter physicists to unravel the deep and intricate features of many-body systems.

\ack
This work was financially supported by Conacyt grant 287067.

%

\section*{References}
\bibliographystyle{iopart-num}
\bibliography{references}

\end{document}